\documentclass[12pt,a4paper]{article}
\usepackage[utf8]{inputenc}
\usepackage[english]{babel}
\usepackage{amsmath}
\usepackage{amsfonts}
\usepackage{amssymb}
\usepackage{graphicx}
\usepackage{cancel}
\usepackage{color}
\usepackage{slashed, epsf, latexsym}
\usepackage{epsfig}
\usepackage{graphics}
\usepackage{jheppub}

\begin{document}
\preprint{}
	\title{ Fluid-gravity and membrane-gravity dualities - Comparison at subleading orders}
	\author[]{Sayantani Bhattacharyya} 
	\author[]{, Parthajit Biswas}
	\author[]{, Anirban Dinda}
	\author[]{and Milan Patra}
	\affiliation[]{National Institute of Science Education and Research, HBNI, Bhubaneshwar 752050, Odisha, India}
	\emailAdd{sayanta@niser.ac.in}
	\emailAdd{parthajit.biswas@niser.ac.in}
	\emailAdd{anirban.dinda@niser.ac.in}
	\emailAdd{milan.patra@niser.ac.in}

\abstract{In this note we have compared two different perturbation techniques that could be used to generate solutions of Einstein's equations in presence of negative cosmological constant. One of these two  methods is derivative expansion and the other is an expansion in inverse powers of dimension. Both the techniques generate space-time with a singularity shielded by a dynamical event horizon. We have  shown  that in the appropriate regime of parameter space and with appropriate choice of coordinates, the metrics and corresponding horizon dynamics, generated by these two different techniques,  are exactly equal to the order the solutions are known both sides. 
This work is essentially extension of \cite{prevwork} where the authors have shown the equivalence of the two techniques upto the first non-trivial order.}

\maketitle

\section{Introduction}\label{sec:Intro}
Classical evolution of the space-time is governed by Einstein's equations, which are a set of nonlinear partial differential equations. Till date it has been  impossible to solve these equations in full generality, particularly when the geometry has nontrivial dynamics.  In such situations, if we want to handle the problem analytically,  perturbation becomes the most useful tool.  There exist more than one perturbation techniques, that could be used for the case of gravity. Needless to say, to gain a clear understanding for the space of solutions of Einstein's equations, we also have to chart out the interconnections between different  perturbation schemes, available so far.

In this note, we shall  compare  two  perturbation techniques, developed to handle both the nonlinearity and the dynamics in Einstein's equations in presence of negative cosmological constant, namely `derivative expansion' \cite{nonlinfluid,arbitdim,Hubeny:2011hd} and `large-$D$ expansion' \cite{membrane,Chmembrane,yogesh1,arbBack,secondorder}\footnote{see \cite{Emparan:2013moa,Emparan:2014cia,Emparan:2015hwa,EmparanHydro,Emparan:2015gva,Tanabe:2015hda,yogesh2,chargepoulomi,Dandekar:2017aiv,Chen:2015fuf,BinChen:2018apr,Chen:2017rxa,Chen:2017hwm,Chen:2017wpf,Chen:2016fuy,Herzog:2017my,Herzog:2016hob,Mandlik2018,Saha2019,Sadhu:2018zyh,Sadhu:2018asi} for work related to `large-$D$ expansion', see \cite{Haack:2008cp,Bhattacharyya:2014bha,Bhattacharyya:2016xfs,Bhattacharyya:2013lha,Bhattacharyya:2007vs,Banerjee:2012iz,Banerjee:2008th,Bhattacharyya:2008ji,krausLarsen} for work related to `derivative expansion'}. The initial set-up for such calculation has already been worked out in \cite{prevwork} along with a comparison at the first non-trivial order(i.e, the leading and the first subleading) on both sides . Here we shall essentially extend it to the second subleading order. Very briefly, what we have done is to re express the metric dual to second order hydrodynamics \cite{arbitdim} , derived using `derivative expansion technique' in the form of the metric dual to membrane dynamics\cite{secondorder} derived using `large-$D$ expansion technique'  upto corrections of order ${\cal O}\left(1\over \text{Dimension}\right)^3$.
As one might have expected, at this order the comparison and matching of the two gravity solutions in the regime of overlap, become far more non trivial than what has been done in \cite{prevwork}.\\

In the next subsection we shall very briefly sketch the strategy we have used for comparison. In fact we shall only give a sketch of the algorithm and shall refer to \cite{prevwork} for any proof or other logical details.

\subsection{Strategy}\label{sec:strategy}
The `large-$D$ expansion' is a technique to generate perturbative gravity solution expanded around space-time dimension $D \rightarrow\infty$ in inverse power of $D$. The metric ${\cal W}_{AB}$, constructed using this method, always has a  `split' form between background $\bar {\cal W}_{AB}$ and rest $ {\cal W}^{(rest)}_{AB}$. $\bar {\cal W}_{AB}$ is the metric of the asymptotic geometry, which is also an exact solution of the Einstein's equations. For our case it is just the  pure AdS.\\
 The classifying data for different ${\cal W}_{AB}$ is encoded by the shape of  a co-dimension one  dynamical hypersurface embedded in pure ADS, coupled with a velocity field. We shall denote the equation that governs the dynamics of this membrane and the velocity field as `membrane equation'. For every solution of this `membrane equation' the `large-$D$ expansion' technique generates one unique dynamical metric that solves Einstein's equations in presence of negative cosmological constant.\footnote{The presence of cosmological constant is not a must for the applicability of the `large-$D$ expansion' technique, but it has to be present for the other technique, namely `derivative expansion' to work . Since our goal is to compare the solutions generated by these two perturbative techniques, we have to deal with Einstein's equations in presence of cosmological constant for both the cases.}
 
 The  technique of `derivative expansion' generates gravity solutions in $D$ dimension that are dual to $(D-1)$ dimensional dynamical fluids, i.e., the characterizing data of the solution is given by a $(D-1)$ dimensional fluid velocity  and temperature field. The velocity and the temperature are assumed to be slowly varying functions of the $(D-1)$ dimensional space . Therefore, the derivatives of these fields are the small parameters  that control the perturbation here.  The dynamics of the fluids are governed by a relativistic (and also higher order) generalization of Navier-Stokes equations, which we shall refer to as `fluid equation'. The duality states that for every solution to the fluid equation, there exists a solution to Einstein's equations in presence of negative cosmological constant, constructed in derivative expansion. For convenience, we shall refer to this metric as `hydrodynamic metric'. This technique works in any  number of space-time dimension. Also note that the metric here is not in a `split form' as we have in the case of `large-$D$' expansion.

In \cite{prevwork}  it has been argued that there exists an overlap in the allowed parameter regimes where these two perturbation techniques  are applicable and also the starting points for both of these  techniques (i.e., the solution at zeroth order) could be chosen to be the same space-time -  namely the black-brane. Now given the zeroth order solution, both `large-$D$' and `derivative expansion' technique generates the higher order solutions uniquely in terms of the characterizing data. Hence it follows that in the overlap regime the two metrics generated by these two techniques must be same or at least coordinate equivalent to each other.\\
Our goal is to show this equivalence in this overlap regime in the space of the perturbation parameters.
We shall do it in the following three key steps.

\subsubsection{Part-1:}
As mentioned above, the metric generated in the `large-$D$' expansion technique would always be expressed as a sum of two metrics - the background $\bar{\cal W}_{AB}$ and ${\cal W}_{AB}^{(rest)}$. The split is such that the contraction of a certain null geodesic vector $O^A\partial_A$ with  ${\cal W}_{AB}^{(rest)}$ vanishes to all order. However, the hydrodynamic metric, to begin with, does not have this `split' form.

Our first step is to split the hydrodynamic metric into `background ' and `rest' such that the background is a pure AdS (though would have a complicated look if we stick to the coordinate system used in \cite{arbitdim}) and the `rest' part of the  metric is such that its contraction with a certain null geodesic vector always vanishes.\\
The procedure is as follows.
\begin{enumerate}
\item We determine the position of the horizon (in an expansion in terms of the  derivatives of  the fluid data) in the hydrodynamic metric following the method described in \cite{entropycurrent}.
\item Next we determine a null geodesic field (affinely parametrized) $\bar O^A\partial_A$, that passes through the horizon.\\
Because of  the specific gauge of the hydrodynamic metric, we could guess a simple form for $\bar O^A\partial_A$ that would work to all order in derivative expansion. We gave some heuristic argument in support of this all order statement.

\item Next we pick up a coordinate system denoted as $\{Y^A\}\equiv\{\rho, y^\mu\}$ such that the `background' of the hydrodynamic metric takes the following form
\begin{equation}\label{eq:backsimp}
ds_\text{background}^2 = \bar G_{AB}~ dY^A dY^B ={d\rho^2\over \rho^2} + \rho^2\eta_{\mu\nu}~ dy^\mu~dy^\nu
\end{equation}
The $\{Y^A\}$ coordinates are related to the $\{X^A\}$ coordinates (the coordinates used in \cite{arbitdim} to express the hydrodynamic metric) by some (yet unknown) mapping functions $f^A(X)$.
$$Y^A = f^A(X)$$

\item Now we demand that the following set of equations.
\begin{equation}\label{eq:mapdef}
\begin{split}
 \bar O^A ~ {\cal G}_{AB}\vert_{\{X\}}=  \bar O^A\left( {\partial f^C\over \partial X^A}\right)\left( {\partial f^{C'}\over \partial X^B}\right)\bar{G}_{CC'}\vert_{\{X\}}
\end{split}
\end{equation}
Where, ${\cal G}_{AB}$ is the full hydrodynamic metric in $\{X^A\}$ coordinates. 
Here the subscript $\{X\}$ denotes that both LHS and RHS of the above equation has been expressed in terms of $\{X\}$ variables.

\item Solving equation \eqref{eq:mapdef} we determine the mapping functions  $f^A$ s. \\
However, it turns out that equation \eqref{eq:mapdef} cannot fix  $f^A$ s uniquely.  To  fix this ambiguity we demanded some extra `conformal type'  symmetry  (see section (\ref{sec:hydsplit}) for the details) on the background metric .\\
As with the case of null geodesic $\bar O^A\partial_A$,  here also we try to guess  some `all order 'expressions  for the mapping functions.

\item Once we know the mapping functions, it is not difficult to see the split of the hydrodynamic metric.
\item Finally we take the large - $D$ limit of the hydrodynamic metric written in a `split'  form.
Our goal is to match this metric with the large-$D$ metric as determined in \cite{secondorder} after expressing the later in terms of fluid - data.
\end{enumerate}
Note all but the  last step in this part has been done exactly in $D$. We have also tried to make some `all order statements'  in terms of derivative expansion, whenever possible.

\subsubsection{Part-2:}
Next comes  the relation between the data of the `large - $D$' expansion technique and that of the derivative expansion. The metric generated in large - $D$ expansion is expressed in terms of a very specific function $\psi$ and the geodesic form field $O_A$, which is not affinely parametrized. It turns out that this $O_A$ is related to the dual form field $\bar O_A$, determined in the previous part (which was affinely parametrized by construction), by an overall normalization. The normalization crucially depends on the  shape of the constant $\psi$ hypersurfaces.\\
So in the second part we first determine $\psi$ and then the normalization of $O^A\partial_A$  in terms of the `fluid data'. The steps are as follows.
\begin{enumerate}
\item According  \cite{secondorder} the function $\psi$ is such that $\psi^{-D}$ is a harmonic function in the   embedding space of the background and also $\psi =1$ is the hypersurface given by the equation of the horizon.
\begin{equation}\label{eq:condpsi_0}
\begin{split}
&\nabla^2\psi^{-D} =0,~~~\text{where $\nabla\equiv$ covariant derivative w.r.t background} \\
&\text{Equation of the bulk horizon}:~ \psi =1\\
\end{split}
\end{equation}
Since we already know the explicit form of the background geometry, the above condition could be solved exactly in $D$ using derivative expansion.
\item The null geodesic $O^A$ is normalized such that 
\begin{equation}\label{condNormO}
\begin{split}
&O^An_A =1~~\text{everywhere in the background}\\
&\text{where $n_A$ is the unit normal to constant $\psi$ hypersurfaces}
\end{split}
\end{equation}
Now we already know the expressions for $\bar O^A$, which is proportional to $O^A$, the required geodesic. Let us denote the proportionality constant as $\Phi$.
\begin{equation}\label{eq:Phi}
\bar O^A = \Phi~ O^A,~\Rightarrow~
\Phi =n_A\bar O^A
\end{equation}
Clearly once we know both $\psi$ and $\bar O^A$, it is easy to determine $\Phi$ and therefore $O^A$ in terms of fluid data, all are exact in $D$. 
\item We substitute these expressions of $\psi$ and $O^A\partial_A$ in the `large-$D$' metric as derived in \cite{secondorder} and convert it to metric in terms of fluid data. 
\end{enumerate}
After following the above steps in this part, we find a metric which we expect to match with the metric found in the previous part upto appropriate orders in derivative and $\left(1\over D\right)$ expansion.

\subsubsection{Part-3:}
As mentioned before, in case of `large-$D$' expansion, the characterizing data of the metric  consist of the shape  of $\psi =1$ membrane viewed as a hypersurface embedded in the background pure AdS and a coupled $D-1$ dimensional  velocity field  (we shall refer to this data set as `membrane data'). In `derivative expansion' the data are the velocity and temperature of a relativistic fluid living on a $(D-1)$ dimensional Minkowski space (referred to as `fluid data'). 

But for both the cases, we are not allowed to choose these data completely arbitrarily; they are constrained by some equations.  For `large-D' expansion this is the membrane equation that governs the coupled dynamics of the membrane shape and the velocity. For `derivative expansion' it is simply the relativistic generalization of the Navier Stokes equation.

After completing the previous two parts, we would be able to identify the membrane data in terms of the fluid data. However, as we shall see in the later sections, this identification will be done locally point by point,  both in time and space. If we want the relations  between these two sets of data to be   valid always and everywhere, then their dynamics must be compatible. In other words, if we rewrite the membrane equation in terms of the fluid data it should reduce to `relativistic Navier Stokes equation' in the appropriate limit of large dimension.\\

The fluid equation (the governing equation for fluid data) could be expressed as conservation of a specific  stress tensor  $ T_{\mu\nu}$ living  on a flat $(D-1)$ dimensional space-time .
\begin{equation} \label{eq:fluidconst}
\begin{split}
&\partial_\mu T^\mu_\nu =0,\\
\end{split}
\end{equation}
In \cite{radiation} the authors have expressed the membrane equation also in terms of a stress tensor  $\hat T_{ab}$ living on $\psi =1$ hypersurface and  conserved with respect to the induced metric of membrane upto correction of ${\cal O}\left(1\over D\right)^2$ .
The membrane equation (the governing equation for large-$D$ data) could be expressed as 
\begin{equation} \label{eq:largeDconst}
\begin{split}
&\tilde\nabla_a \hat T^a_b =0,\\
&\tilde\nabla_a = \text{covariant derivative w.r.t the induced metric of the membrane}
\end{split}
\end{equation}
It turns out that  the existence  of $\hat T_{ab}$ makes the comparison quite easy.
We took the following steps.
\begin{enumerate}
\item To begin with $\hat T_{ab}$ is a function of the membrane data encoded in the extrinsic curvature of the $\psi=1$ hypersurface (the horizon in the bulk geometry) and the velocity field of the membrane, read off from the horizon generators.\\
Fluid stress tensor $T_{\mu\nu}$ is a function of fluid velocity and the temperature.

\item But we already know the precise form of horizon generator and  the $\psi =1$ hypersurface in terms of fluid data. Therefore we could easily compute the extrinsic curvature of the surface as well as the induced metric on it in terms of the coordinates of the flat Minkowski space-time.

\item Inserting these relations in the membrane equation \eqref{eq:largeDconst}, we first convert the equation in form
$\partial_\mu W^{\mu\nu}=0$ for some tensor $W^{\mu\nu}$
\item Finally we match $W^{\mu\nu}$ with $T^{\mu\nu}$ upto the appropriate order in large-$D$ and derivative expansion.
\end{enumerate}
Unfortunately the expression for $\hat T_{ab}$ is not known at order ${\cal O}\left(1\over D\right)^2$ though we know the form of membrane equation at that order\cite{secondorder}. So at that order we had to deal with the full membrane equation and showed the equivalence with the help of Mathematica.

\vspace{.5cm}
This note is organized as follows.\\
In section-(\ref{sec:hydmet}) and section-(\ref{sec:largeDmetric}) we simply quote the hydrodynamic and the `large-$D$' metric along with the corresponding constraint equations from \cite{arbitdim} and \cite{secondorder} respectively. Next in sections-(\ref{sec:hydsplit}), (\ref{sec:largeDfluid}) and (\ref{sec:equieq}) we implement the strategy we described in subsection-(\ref{sec:strategy}). Finally in section-(\ref{sec:conclude}) we summarize our work and discussed the future directions.\\
 At this statge we should emphasize that though, in principle, the strategy used in this paper is very similar to \cite{prevwork}, it differs a lot in details. We believe that now we have a more streamlined and simplified procedure to implement the strategy, mentioned in the previous section. However, to establish a clear connection with \cite{prevwork} we have also worked out every details by  following \cite{prevwork} exactly and we have presented this method of work in appendix-(\ref{app:followprev}). \\
Various computational details are collected in the appendices -(\ref{app:largeD}), (\ref{app:horizon}) and (\ref{app:identities}). In appendix-(\ref{app:notation}) we summarized the notations used in this note.

\section{Hydrodynamic metric and its large $D$ limit}\label{sec:hydmet}
The hydrodynamic metric in arbitrary dimension has been derived in \cite{arbitdim}, correctly upto second order in derivative expansion. In this section we shall simply quote  the final result for the metric, position of the horizon and the dual stress tensor from \cite{arbitdim}.

\subsection{Hydrodynamic metric upto 2nd order in derivative expansion}
  The metric dual to  relativistic hydrodynamics in any dimension could be expressed in terms of the basic variables of the dual fluid, living on  $(D-1)$ dimensional flat space. In this case, it is the relativistic velocity, given by the unit normalized four-vector $u^\mu$ and the temperature scale, set by $r_H(x)$ (local fluid temperature is given by the following formula  $T(x)=\left(\frac{D-1}{4\pi}\right) r_H(x)$) .  At $n$th order in derivative expansion the metric has terms with $n$ number of derivatives, acting on $u^\mu(x)$ and $r_H(x)$.\\
   The authors in \cite{arbitdim} have determined the metric corrections for $n=0,1~\text{and}~2$. Independent fluid data at first and second order in derivative expansion are listed in Table-\ref{table:1storder} and Table-\ref{table:2ndorder}. \\
\begin{equation}\label{eq:met}
dS^2 = dS_0^2 + dS_1^2 + dS_2^2
\end{equation}
where,\\
\textbf{Zeroth order Piece:}
\begin{equation}\label{eq:met0}
\begin{split}
dS_0^2 =& -2 u_\mu ~dx^\mu~ dr - r^2 f({\bf r})~u_\mu u_\nu~dx^\mu ~dx^\nu+r^2 {\cal P}_{\mu\nu} dx^\mu dx^\nu\\
&{\cal P}_{\mu\nu}= \eta_{\mu\nu}+u_\mu u_\nu,~~~~{\bf r} = r/r_H,~~~~ f(z)=1-z^{-(D-1)}
\end{split}
\end{equation}
\textbf{First order Piece:}
\begin{equation}\label{eq:met1}
\begin{split}
&dS_1^2 = -r \left(u_\mu A_\nu + u_\nu A_\mu\right) dx^\mu~dx^\nu + 2 r~ F({\bf r})~\sigma_{\mu\nu} ~dx^\mu~dx^\nu\\
&\text{where,}\\
&A_\mu =(u\cdot\partial)u_\mu -\left(\partial\cdot u\over D-2\right) u_\mu,~~~~\sigma_{\mu\nu}={\cal P}_\mu^\alpha {\cal P}_\nu^\beta\left[\frac{\partial_\alpha u_\beta+\partial_\beta u_\alpha}{2}-{\eta}_{\alpha\beta}\left(\frac{\partial\cdot u}{D-2}\right)\right]\\
&{\bf r} = r/r_H,~~~~~~F(y) = y\int_y^\infty {dx\over x}\left[ x^{D-2} -1\over x^{D-1} -1\right]
\end{split}
\end{equation}
\textbf{Second order Piece:}
\begin{equation}\label{eq:met2}
\begin{split}
&dS_2^2 = \bigg[X_1 ~u_\mu u_\nu + X_2 {\cal P}_{\mu\nu} +\left(Y_\mu u_\nu + Y_\nu u_\mu\right) + Z_{\mu\nu}\bigg]dx^\mu~dx^\nu\\
\text{where,}&\\
X_1 = &~-\bigg[2\left(\partial\cdot A\over D-3\right)-A^2  + \omega^2\left({1\over 2{\bf r}^{D-1}} + {2\over D-3}\right)   + {\sigma^2\over D-2}\left\{{K_2({\bf r})\over {\bf r}^{D-3}}-2\left({D-2\over D-3}\right)  \right\}    \bigg]\\
X_2 =&~ \bigg[2~[F({\bf r} )]^2- K_1({\bf r})\bigg] \left(\sigma^2\over D-2\right)+\frac{\omega^2}{D-2}\\
\end{split}
\end{equation}
\begin{equation}
\begin{split}
Y_\mu =& \left[2\over {\bf r}^{D-3}\right]\left[ L({\bf r} )+ {{\bf r}^{D-3}\over 2(D-3)}\right] \left({\cal D}_\lambda\sigma^\lambda_\mu \right) - \left[1\over D-3\right] \left({\cal D}_\lambda{\omega^\lambda}_\mu \right) \\
Z_{\mu\nu} =&~\bigg( 2~[F({\bf r})]^2 - H_1({\bf r})\bigg)\sigma_\mu^\lambda\sigma_{\lambda\nu} -{\cal P}_{\mu\nu}\bigg( 2~[F({\bf r})]^2 - H_1({\bf r})\bigg)\left(\frac{\sigma^2}{D-2}\right)\\
&+\bigg[H_2({\bf r} )- H_1({\bf r})\bigg]  (u\cdot{\cal D})\sigma_{\mu\nu} + H_2({\bf r} )\bigg({\omega_\mu}^\lambda \sigma_{\lambda\nu} +{\omega_\nu}^\lambda \sigma_{\lambda\mu}\bigg)+\left[{\omega_\mu}^\lambda \omega_{\nu\lambda}-{\cal P}_{\mu\nu}\left(\frac{\omega^2}{D-2}\right)\right]
\end{split}
\end{equation}
where,
\begin{equation}\label{eq:notation1}
\begin{split}
&{\bf r} = r/r_H\\
&\omega_{\mu\nu}= {\cal P}_\mu^\alpha {\cal P}_\nu^\beta\left(\partial_\alpha u_\beta -\partial_\beta u_\alpha\over 2\right),~~\sigma^2 = \sigma_{\mu\nu}\sigma^{\mu\nu},~~\omega^2 = \omega_{\mu\nu} \omega^{\mu\nu},\\
&(u\cdot{\cal D})\sigma_{\mu\nu} = {\cal P}_\mu^\alpha {\cal P}_\nu^\beta~ (u\cdot\partial) \sigma_{\alpha\beta} +\left(\Theta\over D-2\right) \sigma_{\mu\nu}\\
&{\cal D}^\lambda\sigma_{\mu\lambda}= {\cal P}_\mu^\alpha~\partial^\lambda\sigma_{\alpha\lambda} - (D-2) A^\lambda~ \sigma_{\mu\lambda}\\
&{\cal D}^\lambda\omega_{\mu\lambda}= {\cal P}_\mu^\alpha~\partial^\lambda~\omega_{\alpha\lambda} - (D-4) A^\lambda \omega_{\mu\lambda}\\
\end{split}
\end{equation}

\begin{equation}\label{eq:notation2}
\begin{split}
&H_1(y) = 2y^2\int_y^\infty {dx\over x}\left[x^{D-3} -1\over x^{D-1} -1\right]\\
&H_2(y) = {F(y)^2}- 2~y^2\int _y ^\infty{dx\over x(x^{D-1}-1)}\int_1^x{dz\over z}\left[z^{D-3}-1\over z^{D-1}-1\right]\\
&K_1(y) = 2y^2\int_y^\infty {dx\over x^2}\int _x^\infty {dz\over z^2}\bigg[ z~ F'(z) - F(z)\bigg]^2\\
&K_2(y)= \int_y^\infty\left(dx\over x^2\right)\bigg[1-2(D-2)~x^{D-2} -\left(1-{1\over x}\right)\bigg(xF'(x) - F(x)\bigg)\\
&~~~~~~~~~~~~~~~~~~~~~~~~~~~~~+\bigg(2(D-2)x^{D-1} - (D-3)\bigg)\int_x^\infty {dz\over z^2}\bigg(zF'(z) - F(z)\bigg)^2\bigg]\\
&L(y)=\int_y^\infty dx~x^{D-2}\int_x^\infty {dz\over z^3}\left[z-1\over z^{D-1} -1\right]
\end{split}
\end{equation}
This is a dynamical black-brane metric with a singularity ar $r=0$ and the location of the horizon is given by
\begin{equation}\label{eq:hor}
\begin{split}
H(x) =&~r_H(x)+\frac{1}{r_H(x)}\bigg[h_1~\sigma^{\mu\nu}\sigma_{\mu\nu}+h_2~\omega^{\mu\nu}\omega_{\mu\nu}\\
&~~~~~~~~~~~~+h_3(D-3)\left\{\left(\frac{\Theta}{D-2}\right)^2-a^2+2~(u\cdot\partial)\left(\frac{\Theta}{D-2}\right)\right\}\bigg]
\end{split}
\end{equation}
Where,
\begin{equation}
\begin{split}
&h_1=\frac{4}{(D-1)^2(D-2)}-\frac{K_{2H}}{(D-1)(D-2)},~~~h_2=-\frac{1}{2(D-1)}~~~\text{and,}~~~ h_3=-\frac{1}{(D-1)(D-3)}\\
&\text{with,}~~~K_{2H}= \int_1^\infty\left(dx\over x^2\right)\bigg[1-2(D-2)~x^{D-2} -\left(1-{1\over x}\right)\bigg(xF'(x) - F(x)\bigg)\\
&~~~~~~~~~~~~~~~~~~~~~~~~~~~~~+\bigg(2(D-2)x^{D-1} - (D-3)\bigg)\int_x^\infty {dz\over z^2}\bigg(zF'(z) - F(z)\bigg)^2\bigg]
\end{split}
\end{equation}
The fluid dual to the metric, described above, is characterized by the following stress tensor, living on $(D-1)$ dimensional flat Minkowski space
\begin{equation}\label{eq:fluidstress_1}
\begin{split}
T_{\mu\nu}=p~[~\eta_{\mu\nu}+(D-1)u_\mu u_\nu]-2~\eta~\sigma_{\mu\nu}
\end{split}
\end{equation}
Where,
\begin{equation}
p=\frac{r_H^{D-1}}{16\pi G_{\text{AdS}}}~~~\text{and,}~~~\eta=\frac{r_H^{D-2}}{16\pi G_{\text{AdS}}}
\end{equation}
The hydrodynamic metric would solve the $D$ dimensional Einstein's equations  in presence of negative cosmological constant  provided the stress tensor described in equation \eqref{eq:fluidstress_1} is conserved.

\begin{table}[ht]
\caption{Data at 1st order in derivative} % title of Table
\vspace{0.5cm}
\centering % used for centering table
\begin{tabular}{|c| c|} % centered columns (4 columns)
\hline\hline %inserts double horizontal lines
 &Independent Data  \\ [1ex] % inserts table
%heading
\hline % inserts single horizontal line
\hline
Scalar & $\frac{\Theta}{D-2}=\left(\frac{\partial\cdot u}{D-2}\right) $\\ [1ex]
\hline
Vector &$a_\mu=(u\cdot\partial)u_\mu$\\ [1ex]
\hline
\vspace{-0.3cm}
& \\
Tensor &$\sigma_{\mu\nu}={\cal P}_\mu^\alpha {\cal P}_\nu^\beta\left[\frac{\partial_\alpha u_\beta+\partial_\beta u_\alpha}{2}-{\eta}_{\alpha\beta}\left(\frac{\Theta}{D-2}\right)\right]$\\ [1ex]
\hline
\hline
\end{tabular}
\label{table:1storder} % is used to refer this table in the text
\end{table}

\begin{table}[ht]
\caption{Data at 2nd order in derivative} % title of Table
\vspace{0.5cm}
\centering % used for centering table
\begin{tabular}{|c| c|} % centered columns (4 columns)
\hline\hline %inserts double horizontal lines
 &Independent Data  \\ [1ex] % inserts table
%heading
\hline % inserts single horizontal line
\hline
\vspace{-0.3cm}
& \\
Scalars & $ \mathfrak{s}_1\equiv\left(\frac{\Theta}{D-2}\right)^2$,~~ $\mathfrak{s}_2\equiv a^2$,~~$\mathfrak{s}_3\equiv \omega^{\mu\nu}\omega_{\mu\nu}$,~~$\mathfrak{s}_4=\sigma^{\mu\nu}\sigma_{\mu\nu}$,~~$\mathfrak{s}_5=(u\cdot\partial)\left(\frac{\Theta}{D-2}\right)$\\ [1ex]
\hline
\vspace{-0.3cm}
& \\
Vectors &$\mathfrak{v}^{(1)}_\mu\equiv\left(\frac{\Theta}{D-2}\right)a_\mu $,~~$\mathfrak{v}^{(2)}_\mu\equiv a^\nu \omega_{\nu\mu}$,~~$\mathfrak{v}^{(3)}_\mu\equiv a^\nu \sigma_{\nu\mu}$,\\[1ex]
&$\mathfrak{v}^{(4)}_\mu\equiv {\cal P}_{\mu\nu}\partial^\nu\left(\frac{\Theta}{D-2}\right)$,~~$\mathfrak{v}^{(5)}_\mu\equiv {\cal P}_{\mu\nu}\left(\frac{\partial_\lambda \sigma^{\nu\lambda}}{D-2}\right)$\\[2ex] 
\hline 
\vspace{-0.3cm}
& \\
Tensors &$\mathfrak{t}^{(1)}_{\mu\nu}\equiv {\sigma_\mu}^\alpha\sigma_{\alpha\nu}$,~~$\mathfrak{t}^{(2)}_{\mu\nu}\equiv{\omega_\mu}^\alpha\omega_{\alpha\nu}$,~~$\mathfrak{t}^{(3)}_{\mu\nu}\equiv{\omega_\mu}^\alpha\sigma_{\alpha\nu}-{\sigma_\mu}^\alpha\omega_{\alpha\nu}$\\ [1ex]
&$\mathfrak{t}^{(4)}_{\mu\nu}\equiv {\cal P}_\mu^\alpha {\cal P}_\nu^\beta(u\cdot\partial)\sigma_{\alpha\beta}$,~~$\mathfrak{t}^{(5)}_{\mu\nu}\equiv a_\mu a_\nu$\\ [1ex]
\hline
\hline
\end{tabular}
\label{table:2ndorder} % is used to refer this table in the text
\end{table}
Where, ${\cal P}_{\mu\nu}={\eta}_{\mu\nu}+u_\mu u_\nu$

\section{Large-$D$ metric and Membrane equation}\label{sec:largeDmetric}
Just like the previous section, here we shall simply quote the form of the large-$D$ metric from \cite{secondorder}, correctly upto order ${\cal O}\left(1\over D\right)^2$. Schematically, the solution generated by Large-$D$ technique takes the following form
\begin{equation}\label{schemesol}
{\cal W}_{AB} = {\cal W}^{(0)}_{AB} +\left(\frac{1}{D}\right) {\cal W}^{(1)}_{AB} +\left(\frac{1}{D}\right)^2 {\cal W}^{(2)}_{AB}  + \cdots
\end{equation}
Where, the starting ansatz ${\cal W}^{(0)}_{AB}$ is given by
\begin{equation}\label{ansatz}
\begin{split}
{\cal W}^{(0)}_{AB} &= \bar{\cal W}_{AB} + \psi^{-D}O_A O_B
\end{split}
\end{equation}

Here $\bar{\cal W}_{AB}$ is the background metric which could be any smooth solution of the Einstein's equations. The function $\psi(X^A)$ and the one-form field $O_A\equiv O_A dX^A$ are defined in section- (\ref{sec:strategy}).
Rest of the metric corrections could be expressed in terms of $O_A$, $\psi$ and their derivatives.\\
 For convenience, one velocity field has been defined on the constant $\psi$ slices as follows.
\begin{equation}\label{eq:defU}
\begin{split}
&U_A =  n_A - O_A\\
&\text{where,}\\
&\text{ $n_A\equiv$  unit normal to constnt $\psi$ hypersurfaces embedded in background.}
\end{split}
\end{equation}
And the derivatives of $O_A$ has been replaced by derivatives of $U_A$ and $n_A$ or the extrinsic curvature of the constant $\psi$ surfaces.

It turns out that ${\cal W}^{(1)}_{AB}$ - 1st order metric correction simply vanishes.\\
${\cal W}^{(2)}_{AB}$- 2nd order metric correction is non-zero. It can be decomposed as follows.
\begin{equation}\label{eq:parameter1}
\begin{split}
&{\cal W}^{(2)}_{AB} =\bigg[O_A O_B\left(\sum_{n=1}^2 f_n(R)~S_n \right)+  v(R)~\big(~ V_A O_B + V_B O_A\big) +  t(R)~T_{AB} 
\bigg]\\
\end{split}
\end{equation}\\
where,
\begin{equation}\label{structurevecten}
\begin{split}
&T_{AB}=P^C_A P^{D}_B\bigg[ \bar{R}_{FCDE}O^E O^F+\frac{K}{D}\bigg(K_{CD}-\frac{\nabla_C U_{D}+\nabla_{D}U_C}{2}\bigg)\\
&~~~~~~~~~~~~~~~~~~- P^{EF}(K_{EC}-\nabla_E U_C)(K_{FD}-\nabla_F U_{D})\bigg]\\
\end{split}
\end{equation}
\begin{equation}
\begin{split}
&V_{A}={P}^B_{A}\Bigg[\frac{K}{D}\left(n^D U^E O^F \bar{R}_{FBDE}\right)+\frac{K^2}{2D^2}\left(\frac{{\nabla}_B {K}}{K}+(U\cdot{\nabla})U_B-2~U^D {K}_{D B}\right)\\
&~~~~~~~~~~~~~-{P}^{F D} \left(\frac{{\nabla}_F {K}}{D}-\frac{K}{D} (U^E {K}_{E F})\right)\left({K}_{D B}-\nabla_D U_B\right)\Bigg]
\end{split}
\end{equation}
\begin{equation}\label{structurescalar}
\begin{split}
&S_1=U^E U^F n^D n^C\bar{R}_{CEFD}+\left(\frac{U\cdot{\nabla}K}{K}\right)^2+\frac{\hat{\nabla}_A {K}}{K}\left[4~U^B {K}^A_B-2\left[(U\cdot{\nabla})U^A\right]-\frac{\hat{\nabla}^A {K}}{K}\right]\\
&-(\hat{\nabla}_A U_B)(\hat{\nabla}^A U^B)-(U\cdot {K}\cdot U)^2-\left[(U\cdot\hat{\nabla})U_A\right][(U\cdot\hat{\nabla})U^A]+2\left[(U\cdot{\nabla})U^A\right](U^B {K}_{BA})\\
&~~~~~~~~~~~~~~-3~(U\cdot {K}\cdot {K}\cdot U)-\frac{K}{D}\left(\frac{U\cdot{\nabla}{K}}{K}-U\cdot {K}\cdot U\right)\\
&S_2=\frac{K^2}{D^2}\Bigg[-\frac{{K}}{D}\left(\frac{U\cdot{\nabla}{K}}{K}-U\cdot {K}\cdot U\right)- 2~\lambda- (U\cdot {K}\cdot {K}\cdot U)+2 \left(\frac{{\nabla}_A{K}}{K}\right)U^B {K}^A_B-\left(\frac{U\cdot{\nabla}{K}}{K}\right)^2\\
&+2\left(\frac{U\cdot{\nabla}{K}}{K}\right)(U\cdot {K}\cdot U)-\left(\frac{\hat{\nabla}^D K}{K}\right)\left(\frac{\hat{\nabla}_D K}{K}\right)-(U\cdot {K}\cdot U)^2+n^B n^D U^E U^F\bar{R}_{FBDE}\Bigg]
\end{split}
\end{equation}
$\bar{R}_{ABCD}$ is the Riemann tensor of the background metric $\bar{\cal W}_{AB}$.\\
$\nabla$ denotes the covariant derivative with respect to $\bar{\cal W}_{AB}$.
$\hat{\nabla}$ is defined as follows:
 for any general  tensor with $n$ indices $W_{A_1A_2\cdots A_n}$
\begin{equation}\label{tildedef}
\hat{\nabla}_A W_{A_1A_2\cdots A_n}=\Pi_A^C~\Pi_{A_1}^{C_1}\Pi_{A_2}^{C_2}\cdots \Pi_{A_n}^{C_n}\left(\nabla_C W_{C_1C_2\cdots C_n}\right), \quad \text{with}\quad \Pi_{AB}=\bar{\cal W}_{AB}-n_A n_B
\end{equation}\\
and,
\begin{equation}\label{funcn}
\begin{split}
&t(R)=-~2\left(\frac{D}{K}\right)^2\int_R^{\infty}\frac{y~dy}{e^y-1}\\
&v(R)=2\left(\frac{D}{K}\right)^3\bigg[\int_R^{\infty}e^{-x}dx\int_0^x\frac{y~e^y}{e^y-1}dy~-~e^{-R}\int_0^{\infty}e^{-x}dx\int_0^x\frac{y~e^y}{e^y-1}dy\bigg]\\
&f_1(R)=-2\left(\frac{D}{K}\right)^2\int_R^{\infty}x~e^{-x}dx+2~e^{-R}\left(\frac{D}{K}\right)^2\int_0^{\infty}x~e^{-x}dx\\
&f_2(R)=\left(\frac{D}{K}\right)\Bigg[\int_R^{\infty}e^{-x}dx\int_0^x\frac{v(y)}{1-e^{-y}}dy-e^{-R}\int_0^{\infty}e^{-x}dx\int_0^x\frac{v(y)}{1-e^{-y}}dy\Bigg]\\
&~~~~~~~~-\left(\frac{D}{K}\right)^4\Bigg[\int_R^{\infty}e^{-x} dx\int_0^x\frac{y^2~ e^{-y}}{1-e^{-y}}dy-e^{-R}\int_0^{\infty}e^{-x} dx\int_0^x\frac{y^2 ~e^{-y}}{1-e^{-y}}dy\Bigg]\\
&\text{Where,~~~~}R\equiv D(\psi-1)
\end{split}
\end{equation}\\

The above expressions for ${\cal W}_{AB}$ would solve Einstein's equations in presence of negative cosmological constant\footnote{Note that each component of the metric corrections described above vanishes exponentially in $D$  when $R\sim{\cal O}(D)$. Now this `large -$D$ metric', by construction, solves  Einstein's equations (in presence of negative cosmological constant) upto correction of order ${\cal O}\left(1\over D\right)^3$; and therefore, whenever the metric corrections become of the order of ${\cal O}(e^{-D})$, they are no longer trustable. In other words, the above metric solves Einstein's equations as long as $R= D(\psi-1)<<D$.\\
 It follows that while comparing with hydrodynamic metric we would expect a perfect match only within this region of validity of the large-$D$ metric. Also a `match' requires a similar exponential fall off in $D$ for the hydrodynamic metric if one goes away distance of order ${\cal O}(D)$ from the horizon - the $\psi=1$ hypersurface.} provided the following constraint equation is satisfied,

\begin{equation}\label{eq:constraint2}
\begin{split}
&P^A_C\bigg[\frac{\hat{\nabla}^2 U_A}{{K}}-\frac{\hat{\nabla}_A{{K}}}{{K}}+U^B {{K}}_{B A}-U\cdot\hat{\nabla} U_A\bigg]+P^A_C\bigg[-\frac{U^B {{K}}_{B D} {{K}}^D_A}{{K}}+\frac{\hat{\nabla}^2\hat{\nabla}^2 U_A}{{{K}}^3}-\frac{(\hat{\nabla}_A{{K}})(U\cdot\hat{\nabla}{{K}})}{{{K}}^3}\\
&~~-\frac{(\hat{\nabla}_B{{K}})(\hat{\nabla}^B U_A)}{{{K}}^2}-\frac{2{{K}}^{D E}\hat{\nabla}_D\hat{\nabla}_E U_A}{K^2}-\frac{\hat{\nabla}_A\hat{\nabla}^2{{K}}}{{{K}}^3}+\frac{\hat{\nabla}_A({{K}}_{BD} {{K}}^{BD} {{K}})}{K^3}+3\frac{(U\cdot {{K}}\cdot U)(U\cdot\hat{\nabla} U_A)}{{{K}}}
\\
&~~-3\frac{(U\cdot {{K}}\cdot U)(U^B {{K}}_{B A})}{{{K}}}-6\frac{(U\cdot\hat{\nabla}{{K}})(U\cdot\hat{\nabla} U_A)}{{{K}}^2}+6\frac{(U\cdot\hat{\nabla}{{K}})(U^B {{K}}_{B A})}{{{K}}^2}+3\frac{U\cdot\hat{\nabla} U_A}{D-3}\\
&~~-3\frac{U^B {{K}}_{B A}}{D-3}-\frac{(D-1)\lambda}{{{K}}^2}\bigg(\frac{\hat{\nabla}_A {{K}}}{{K}}-2U^D {{K}}_{D A}+2(U\cdot\hat{\nabla})U_A\bigg)\bigg]={\cal O}\left(\frac{1}{D}\right)^2\\ \\
&{\text{and,}}~~~~\hat{\nabla}\cdot U-\frac{1}{2K}\nabla_{(A}U_{B)}\nabla_{(C}U_{D)}P^{AC} P^{BD}={\cal O}\left(\frac{1}{D}\right)^2
\end{split}
\end{equation}
where,~~~$U_A=n_A-O_A$,~~~ $P_{AB}=\bar{\cal W}_{AB}-n_A n_B+U_A U_B$~ and~~$\nabla_{(A}U_{B)}=\nabla_{A}U_{B}+\nabla_{B}U_{A}$\\

If we truncate the membrane equation at first subleading order, it takes the following form
\begin{equation}
\begin{split}
P^A_C\bigg[\frac{\hat{\nabla}^2 U_A}{{K}}-\frac{\hat{\nabla}_A{{K}}}{{K}}+U^D {{K}}_{D A}-(U\cdot\hat{\nabla}) U_A\bigg]= {\cal O}\left(\frac{1}{D}\right),~~\hat{\nabla}\cdot U={\cal{O}}\left(\frac{1}{D}\right)
\end{split}
\end{equation}
In \cite{radiation} this part of the equation has been expressed as a conservation of some stress tensor, defined on the $\psi=1$ hypersurface. The form this stress tensor is as follows.
\begin{equation}\label{eq:memstress}
\begin{split}
T_{AB}^{(m)}&=\left(\frac{ K}{2}\right)U_A U_B+\left(\frac{1}{2}\right){K}_{AB}-\frac{1}{2}\left(\hat{\nabla}_A U_B+\hat{\nabla}_B U_A\right)-\frac{1}{K}\left(U_A\hat{\nabla}^2U_B+U_B\hat{\nabla}^2U_A\right)\\
&~~~~+\frac{1}{2}\left(U_A\frac{\hat{\nabla}_B K}{K}+U_B\frac{\hat{\nabla}_A K}{K}\right)-\frac{1}{2}\left(U\cdot{K}\cdot U+\frac{K}{D}\right)\Pi_{AB}
\end{split}
\end{equation}
As explained in section-(\ref{sec:strategy}), we shall use this form of the membrane equation to show equivalence between the two sets of the constraint equations.

\section{Implementing part-1: \\The split of the hydrodynamic metric}\label{sec:hydsplit}
In this section we shall see how to split the hydrodynamic metric as a sum of the background and the rest. The hydrodynamic metric that we shall work with is correct upto second order in derivative expansion and therefore in this section we shall neglect all terms of third order or higher. As we have mentioned before,  all these steps are already executed in \cite{prevwork} accurately upto first order in derivative expansion. Here we shall use the results derived in \cite{prevwork} whenever possible. Also we shall try to generalize the results and the derivation, as much as possible, to higher orders on both sides of the perturbation. It turns out that often some general pattern emerges which would naturally lead to some `all-order' statements at the intermediate steps.

\subsection{The null geodesic $ \bar O^A\partial_A$}
As summarized in the introduction, the `split ' of the metric would be done in terms of a geodesic field $O^A\partial_A$ which is null with respect to the full space-time and also with respect to the background. In this subsection our task is to fix this $O_A$ field.

Before getting into any details of this second order calculation let us describe few general features of the hydrodynamic metric ${\cal G}_{AB}$, which would allow us to determine a null vector field that would satisfy the geodesic equation to all order in derivative expansion.\\
 According to the derivation of  \cite{arbitdim}, the coordinates are fixed in a way  such that ${\cal G}_{rr}=0$ and ${\cal G}_{r\mu}=-u_\mu$  to all order in derivative expansion.  In this gauge $\Gamma^r_{rr}$ and $\Gamma^\mu_{rr}$ vanish identically to all order. It follows that  in this metric, any vector of the form  $\left(k^A\partial_A\equiv\zeta(x^\mu)\partial_r \right)$ would be an affinely parametrized null geodesic to all order in derivative expansion as long as the function $\zeta$ depends only on $x^\mu$ .\\
\begin{equation}\label{eq:simpchk}
\begin{split}
&(k^A\bar\nabla_A) k^r=k^r\partial_r k^r+ k^r\Gamma^r_{rr} k^r =0\\
&(k^A\bar\nabla_A) k^\mu= k^r\Gamma^\mu_{rr} k^r =0\\
\end{split}
\end{equation}
Now at zeroth order in derivative expansion we know that $\bar O^A\partial_A $ is simply $\partial_r$. In fact this turns out to be true even at first order in derivative expansion\cite{prevwork}. It is very tempting to conjecture that  to all order in derivative expansion
\begin{equation}\label{eq:finOa}
\bar O^A\partial_A = \partial_r
\end{equation}
We could simply set the function $\zeta(x^\mu)$ to be one, since anyway we have to normalize $\bar O^A$ further to get the $O^A$ vector field ( see the previous section) that appears in the large-$D$ metric\\
We could construct some inductive proof for this statement. Suppose at some $n$th order in derivative expansion $\bar O^A\partial_A = 
\partial_r$. At $(n+1)$ th order, after setting the norm  to zero and normalization by fixing the coefficient of $\partial_r$ to be one,  the  form of $\bar O^A$ would be 
$$\bar O^A\partial_A = \partial_r + V^\mu (r) \partial_\mu$$ where $V^\mu$ is some vector structure, perpendicular to $u^\mu$ and  containing $(n+1)$ derivatives.   Now since $ V^\mu (r)$ already contains $(n+1)$ derivatives, in the geodesic equation at $(n+1)$ th order, it is the zeroth order metric that will multiply this term and we could solve for the $r$ dependence of $ V^\mu (r)$ without any details of the higher order metric correction. $V^\mu(r)$  turns out to be
$$V^\mu(r) = {\tilde V^\mu\over r^2},~~~\text{where $\tilde V^\mu$ is independent of $r$}$$
Upon lowering the index we find $\bar O_A ~dX^A =\left[ -u_\mu  + \tilde V_\mu + {\cal O}\left(\partial^{n+2}\right)\right]dx^\mu $.
Substituting this $\bar O_A$ in the expression of large-$D$ metric and using the facts that $\psi^{-D}=  \left(r_H\over r\right)^{D-1} + {\cal O}\left(1\over D\right)$ and $\bar O_A$ is proportional to $O_A$, we could see that  the leading term in $\left(1\over D\right)$ expansion (i.e., the terms $\psi^{-D} O_A O_B$) itself will generate a term of the form $\sim \left(r_H\over r\right)^{D-1}( u_\mu \tilde V_\nu + u_\nu \tilde V_\mu)$. Using AdS-CFT correspondence one could deduce that such a term in the metric will generate a term of the form $( u_\mu \tilde V_\nu + u_\nu \tilde V_\mu)$ in the dual fluid stress tensor, thus making it out of Landau frame. But since $u_\mu$ of the hydrodynamic metric is defined to be the fluid velocity in  Landau frame (see \cite{arbitdim},\cite{nonlinfluid}), such a term in $O^A$ must vanish once we equate the resultant large-$D$ metric with the hydrodynamic metric. \\

Hence equation \eqref{eq:finOa} gives an all order expression for $\bar O^A\partial_A$

\subsection{The mapping functions and the `split'  of the hydrodynamic metric}\label{subsec:map}
Next we come to the computation of the mapping functions $f^A$ s that relate the $\{Y^A\} = \{\rho, y^\mu\}$ coordinates (where the background pure AdS has simple metric given by equation \eqref{eq:backsimp}) with the $\{X^A\}= \{r,x^\mu\}$, the coordinates in which the  hydrodynamic metric ${\cal G}_{AB}$ is expressed in section (\ref{sec:hydmet}).

As before we shall start with some general observation and try to get some all order statements about the mapping functions. We shall use equation \eqref{eq:finOa} for the expression of $\bar O^A$. Now the mapping functions $f^A$ s are determined by solving equation \eqref{eq:mapdef}. We could view the RHS of equation \eqref{eq:mapdef} as pure AdS expressed in $\{X^A\}$ coordinates and contracted with $\bar O^A$. Let us rewrite equation \eqref{eq:mapdef} in this language.\\
Suppose $\bar {\cal G}_{AB}$ denotes the pure AdS metric in $\{ X^A\}$ coordinates, i.e.,
\begin{equation}\label{eq:barg}
\begin{split}
&\bar {\cal G}_{AB} = \left( {\partial f^C\over \partial X^A}\right)\left( {\partial f^{C'}\over \partial X^B}\right)\bar{G}_{CC'}\vert_{\{X\}}\\
&\text{where $\bar{G}_{CC'}$ is given in equation \eqref{eq:backsimp}}
\end{split}
\end{equation}
After using the fact that $\bar O^A\partial_A= \partial_r$, equation \eqref{eq:mapdef} simply implies
\begin{equation}\label{eq:condsimp}
\bar O^A\left({\cal G}_{AB} -\bar {\cal G}_{AB}\right) \equiv\bar O^A{\cal G}^\text{rest}_{AB} =0~\Rightarrow~ {\cal G}^\text{rest}_{rB}=0
\end{equation}
Now we know that the hydrodynamic metric as presented in \cite{arbitdim} is in a gauge where, to all orders in derivative expansion, 
$${\cal G}_{rr}=0,~~{\cal G}_{r\mu}=-u_\mu$$ 
Clearly equation \eqref{eq:condsimp} could be satisfied provided $\bar {\cal G}_{AB}$ is also in the same gauge. In other words, $f^A$s should be such that it transforms the pure AdS metric in a gauge where the $(r\mu)$ component is equal to minus of $u_\mu$ as read off from the hydrodynamic metric and the $(rr)$ component simply vanishes. \\
Note that in any general metric the above condition does not fix the gauge completely; we are left with a residual coordinate transformation symmetry within the $x^\mu$ coordinates. For example, consider the following set of mapping functions.
\begin{equation}\label{eq:allordermap_0}
\begin{split}
\rho = r + \chi(x),~~~y^\mu = x^\mu + {u^\mu \over r +\chi(x)} + \xi^\mu(x)
\end{split}
\end{equation}
The above transformation will take the pure AdS metric to the required gauge (i.e., $\bar{\cal G}_{rr} =0$ and $\bar{\cal G}_{r\mu} =-u_\mu$) to all order in derivative expansion, as along as the function $\chi$ is independent of the $r$ coordinate and $\xi^\mu(x)$ is an arbitrary four-vector, independent of $r$,  satisfying,
\begin{equation}\label{eq:ccoo}
u_\mu\left({\partial\xi^\mu\over \partial x^\nu}\right)=0
\end{equation}
If we demand an exact match between the large-$D$ and hydrodynamic metric, the mapping functions must be in terms of the fluid data. In other words $\chi(x)$ and $\xi^\mu(x)$ must be functions of  $u^\mu$, $r_H$ and their derivatives. On top of that $\xi^\mu(x)$, once expressed in terms  of independent fluid data, is further constrained to satisfy equation \eqref{eq:ccoo} as an identity. Now we would like to show that only solution to \eqref{eq:ccoo} is
\begin{equation}\label{eq:ccsol}
\xi^\mu = c~ u^\mu ,~~~c = \text{constant}
\end{equation}
Suppose $$\xi^\mu = C(x)~ u^\mu + \xi^\mu_\perp(x)~~\text{such that}~~u_\mu\xi^\mu_\perp =0$$

Substituting this expression of $\xi^\mu$ in \eqref{eq:ccoo} we find
\begin{equation}\label{eq:grad}
\begin{split}
&~~~~ u_\alpha\partial_\nu\left[C(x)u^\alpha+\xi^\alpha_\perp(x)\right]=0\\
\Rightarrow~~  &-\partial_\nu C(x)-\xi^\alpha_\perp(x)\partial_\nu u_\alpha=0\\
\end{split}
\end{equation}
$\xi^\alpha_\perp(x)\partial_\nu u_\alpha$ is a gradient function and therefore must satisfy the following integrability condition. 
\begin{equation}\label{eq:integ}
\begin{split}
&~~~~\partial_\mu(\xi^\alpha_\perp\partial_\nu u_\alpha)-\partial_\nu(\xi^\alpha_\perp\partial_\mu u_\alpha)=0\\
\Rightarrow~~&~~~~(\partial_\mu\xi^\alpha_\perp)(\partial_\nu u_\alpha)-(\partial_\nu\xi^\alpha_\perp)(\partial_\mu u_\alpha)=0
\end{split}
\end{equation}
Dotted with $u_\mu$, 
\begin{equation}\label{eq:sss}
\begin{split}
&[(u\cdot\partial)\xi^\alpha_\perp](\partial_\nu u_\alpha)-a_\alpha(\partial_\nu\xi^\alpha_\perp)=0\\
\Rightarrow~~&\left[\partial_\beta \xi^\alpha_\perp\right]\left[-a_\alpha{\cal P}^\beta_\nu+\sigma_{\nu\alpha}u^\beta+\omega_{\nu\alpha}u^\beta+\left(\frac{\Theta}{D-2}\right){\cal P}_{\nu\alpha}u^\beta\right]=0\\
\end{split}
\end{equation}
In the above equation the four terms multiplying $\partial_\beta \xi^\alpha_\perp$ are independent fluid data and therefore their linear combination can never vanish identically. The only way to satisfy equation \eqref{eq:sss} is to set $\partial_\beta\xi^\alpha_\perp$ to zero.
\begin{equation}
\begin{split}
&\partial_\beta \xi^\alpha_\perp=0,~~
\Rightarrow~~ \xi^\alpha_\perp=\text{constant}
\end{split}
\end{equation}
Now in the hydrodynamic metric there is no special special vector apart from $u^\mu$, which is not a constant. Therefore, if we want a term by term matching of the `large-$D$' metric (written in $\{X^A\}$ coordinates) with the hydrodynamic metric, $\xi^\mu_\perp$ itself must vanish.
Substituting in equation \eqref{eq:ccoo} we find
\begin{equation}\label{eq:ccoo2}
\begin{split}
&\partial_\nu C(x) = 0~~\Rightarrow ~~C(x) = c = \text{constant}\\
\end{split}
\end{equation}
From the above discussion, it also follows that exact matching of the two metrics (upto the required order) would be possible only for a very specific choice of $\chi(x)$ and the constant $c$ in equation \eqref{eq:ccsol}. Any other choice, apart from  this specific one, would result in a hydrodynamic metric which would not be exactly same, but coordinate-equivalent to the metric presented in section-(\ref{sec:hydmet}). The corresponding coordinate transformation would simply be a $x^\mu$ dependent shift in the $r$ and $x^\mu$ coordinates.

  Note that the constant $c$  could not have any derivative correction and the computation of \cite{prevwork}, which is correct upto first order in derivative expansion, has already told us that  $c$ has to be set to zero for an exact match between these two metrics.
It turns out that if we choose the function $\chi(x) = \left(\partial_\mu u^\mu\over D-2\right)$,  it does cast the pure AdS metric to the required gauge and everything works out as we wanted i.e., upto second order in derivative expansion, both the metrics match term by term without any further coordinate redefinition. So the final form of the mapping functions\footnote{At this stage it is very tempting to conjecture  equation \eqref{eq:allordermap} to be an all order statement for the coordinate transformation
since generically we should have an order ${\cal O}\left(\partial^2\right)$ term in the $\rho$ redefinition, but it does not appear. }
. 
\begin{equation}\label{eq:allordermap}
\begin{split}
\rho = r - \left(\Theta\over D-2\right) + {\cal O}\left(\partial^3\right),~~~y^\mu = x^\mu + {u^\mu \over \rho} ,~~\text{where}~~\Theta\equiv \partial\cdot u
\end{split}
\end{equation}

After imposing this coordinate transformation the final form of the background metric is as follows,
\begin{equation}\label{eq:allorderbacl}
\begin{split}
\bar{\cal G}_{rr} &=~0,~~~~\bar{\cal G}_{r\mu}= -u_\mu\\
%\bar{\cal G}_{\mu\nu} &=~\left( r - {\Theta\over D-2}\right)^2\eta_{\mu\nu} + \left( r - {\Theta\over D-2}\right)(\partial_\mu u_\nu +\partial_\nu u_\mu) 
% +\left(u_\mu\partial_\nu\Theta+u_\nu\partial_\mu\Theta\over D-2\right) +( \partial_\mu u_\alpha)(\partial_\nu u^\alpha)  \\
\bar{\cal G}_{\mu\nu} 
&=r^2 ({\cal P}_{\mu\nu}-u_\mu u_\nu)+2r\left(\frac{\Theta}{D-2}\right)u_\mu u_\nu+r\left[2~\sigma_{\mu\nu}-a_\mu u_\nu-a_\nu u_\mu\right]+\left[\mathfrak{t}^{(1)}_{\mu\nu}-\mathfrak{t}^{(2)}_{\mu\nu}+\mathfrak{t}^{(3)}_{\mu\nu}\right]\\
&+u_\mu\left[\mathfrak{v}_\nu^{(2)}-\mathfrak{v}_\nu^{(3)}+\mathfrak{v}_\nu^{(4)}\right]+u_\nu\left[\mathfrak{v}_\mu^{(2)}-\mathfrak{v}_\mu^{(3)}+\mathfrak{v}_\mu^{(4)}\right]-u_\mu u_\nu\left[\mathfrak{s}_1-\mathfrak{s}_2+2\mathfrak{s}_5\right]
\end{split}
\end{equation}

 Once we know the background in $\{r,x^\mu\}$ coordinates, we could determine ${\cal G}_{AB}^\text{rest}$ by simply subtracting the background from the full hydrodynamic metric ${\cal G}_{AB}$.
${\cal G}_{rr}^\text{rest}$ and ${\cal G}_{r\mu}^\text{rest}$ are zero by construction. The structure of ${\cal G}_{\mu\nu}^\text{rest}$ is a bit complicated. We first decompose it into the scalar vector and the tensor sectors.
\begin{equation}\label{eq:grestmunu}
\boxed{
\begin{split}
{\cal G}_{\mu\nu}^{\text{rest}}={\cal G}^{(1)}_s u_\mu u_\nu+{\cal G}^{(2)}_s {\cal P}_{\mu\nu}+\left({\cal G}^{(v)}_\mu u_\nu+{\cal G}^{(v)}_\nu u_\mu\right)+{\cal G}^{(t)}_{\mu\nu}
\end{split}}
\end{equation}
Where, ${\cal G}^{(1)}_s$, ${\cal G}^{(2)}_s$, ${\cal G}^{(v)}_\mu$ and ${\cal G}^{(t)}_{\mu\nu}$ have the following forms
\begin{equation}\label{eq:grestscal_0}
\begin{split}
{\cal G}^{(1)}_s&
=r^2[1-f(\textbf{r})]-\left(\frac{1}{2~{\bf r}^{D-1}}\right)\mathfrak{s}_3-\frac{1}{D-2}\left(\frac{K_2({\bf r})}{{\bf r}^{D-3}}\right)\mathfrak{s}_4\\
{\cal G}_s^{(2)}&=\frac{1}{D-2}\bigg[2~[F({\bf r} )]^2- K_1({\bf r})-1\bigg]\mathfrak{s}_4
\\
{\cal G}^{(v)}_\mu&=\frac{2(D-2)}{{\bf r}^{D-3}}L({\bf r})\left(\mathfrak{v}_\mu^{(5)}-\mathfrak{v}_\mu^{(3)}\right)\\
{\cal G}^{(t)}_{\mu\nu}&=-2~r[1-F(\textbf{r})]\sigma_{\mu\nu}+\bigg[2~[F({\bf r} )]^2- H_1({\bf r})-1\bigg]\left[\mathfrak{t}^{(1)}_{\mu\nu}-{\cal P}_{\mu\nu}\frac{\mathfrak{s}_4}{D-2}\right]+[H_2({\bf r})-1]~\mathfrak{t}^{(3)}_{\mu\nu}\\
&~~~~+[H_2({\bf r})-H_1({\bf r})]\left[\mathfrak{t}^{(4)}_{\mu\nu}+\left(\frac{\Theta}{D-2}\right)\sigma_{\mu\nu}\right]\\
\end{split}
\end{equation}
See Table-(\ref{table:2ndorder}) for the definition of ${\mathfrak s}_i$, ${\mathfrak v}_\mu^{(i)}$ and ${\mathfrak t}_{\mu\nu}^{(i)}$

\section{Implementing part-2:\\ Large-$D$ metric in terms of fluid data}\label{sec:largeDfluid}
In the previous section we have recast  the hydrodynamic metric, ${\cal G}_{AB}$ as a sum of `background' (which is just pure AdS but looks complicated in the coordinate system where the full hydrodynamic metric has the simple form) and the `rest'.
$${\cal G}_{AB} = \bar{\cal G}_{AB}  + {\cal G}_{AB}^\text{rest}$$
The large-$D$ metric ${\cal W}_{AB}$ (see section (\ref{sec:largeDmetric})) has exactly this form. We shall simply identify $\bar{\cal W}_{AB}$ with $\bar{\cal G}_{AB}$. Next to show that the hydrodynamic is exactly same as the large-$D$ metric in the appropriate regime, we need to match  ${\cal W}_{AB}^\text{rest}$ expanded in terms of boundary derivative upto second order in derivative expansion with ${\cal G}_{AB}^\text{rest} $, expanded in inverse power of dimension upto order ${\cal O}\left(1\over D\right)^2$. In this section our goal is to rewrite ${\cal W}_{AB}^\text{rest}$ in terms of fluid data.\\
As we have explained before, ${\cal W}_{AB}^\text{rest}$ is expressed in terms of a harmonic function $\psi$ and a null geodesic one-form field $O_A$ (normalized so that the component of $O_A$ along the normal to the constant  $\psi$ hypersurfaces is always one). We have already determined  the null geodesic field upto the normalization. Our next task is to determine $\psi$ in terms of the fluid data.

\subsection{Determining $\psi$}
The function $\psi$ is a harmonic function in the background AdS. 
\begin{enumerate}
\item $\psi$ satisfies the following differential equation everywhere on the background.
\begin{equation}\label{eq:condpsi}
\nabla^2\psi^{-D} =0 ~~\text{where $\nabla$ denotes covariant derivative with respect to the background}.
\end{equation}
\item $\psi=1$ hypersurface corresponds to the horizon, viewed as a surface embedded in the background.
More precisely $\psi =1\Rightarrow r-H(x) =0$ where $H(x)$ is the location of the horizon in the hydrodynamic metric as quoted in equation \eqref{eq:hor}.

\end{enumerate}

In this subsection we shall determine  $\psi$ solving the above two conditions.  
We shall do it in two steps.\\
We shall first solve equation \eqref{eq:condpsi} in $\{\rho, y^\mu\}$ coordinates,  because the expression of Laplacian is far simpler in this coordinate system (the background pure AdS metric is just diagonal here) as compared to the $\{X^A\}=\{r, x^\mu\}$ system (the one that has been used to describe the hydrodynamic metric in section (\ref{sec:hydmet})). We shall assume that in $\{Y^A\}$ coordinates, $\psi=1$ hypersurface is given by
\begin{equation}
\psi=1 \Rightarrow r = H(x)\Rightarrow\rho =\rho_H(y)
\end{equation}
Note  that the above condition will provide   only one boundary condition for the differential equation on $\psi$ and 
this is not sufficient to  determine a function uniquely.  We need one  more condition. The other boundary condition is  implicitly given by writing the harmonic function as $\psi^{-D}$.  It implies that at a point which is order ${\cal O}(1)$ distance away (along any arbitrary direction) from the $\psi =1$ hypersurface, this harmonic function falls off exponentially with $D$. Now clearly increasing $\rho$ keeping all other $y^\mu$ coordinates constant is one way to go away from the $\psi=1$ hypersurface and therefore the harmonic function $\psi^{-D}$ must vanish as $\rho$ goes to $\infty$.  This will provide the required boundary condition.

Still for a generic $\rho_H(y)$, it is difficult to solve the equation explicitly even in $\{Y^A\}$ coordinate system where the pure AdS has a simple form.  However, in this case we have two perturbation parameters and we know the solution at leading order in terms of both of them. 
$$\psi^{-D} =\left( \rho\over \rho_H\right)^{-(D-1) }+ {\cal O}\left(\partial\right)= \left( r\over r_H\right)^{-(D-1)} + {\cal O}\left(\partial\right)$$
This is what will help us to solve the equation. We shall use derivative expansion and determine $\psi$ upto second order. As usual at every order in derivative expansion we shall encounter a universal and also simple second order ordinary differential equation in $\rho$ with some source. For explicit solution,  we need  two integration constants. One of them is fixed by the condition that $\psi=1$ is the horizon. As we have explained above, the other boundary condition  we fix by demanding that $\psi^{-D}$ vanishes as $\rho\rightarrow\infty$. At the moment,  we do not require $\left(1\over D\right)$ expansion to solve for $\psi$.

In $\{\rho, y^\mu\}$ coordinates, the form of $\psi$ turns out to be the following
\begin{equation}\label{eq:psirho}
\psi=\left(\frac{\rho}{\rho_H}\right)^{1-\frac{1}{D}}-\frac{(D-1)}{2~D(D+1)\rho_H^2}\left(\frac{\rho}{\rho_H}\right)^{1-\frac{1}{D}}\left[1-\left(\frac{\rho}{\rho_H}\right)^{-2}\right]\left[(D-2)~\mathfrak{t}_1+~\mathfrak{t}_2\right]+{\cal O}(\partial)^3
\end{equation}
where,
\begin{equation}
\begin{split}
&\mathfrak{t}_1=\left(\frac{\partial^\mu\rho_H}{\rho_H}\right)\left(\frac{\partial_\mu\rho_H}{\rho_H}\right),~~~\mathfrak{t}_2=\left(\frac{\partial^\mu\partial_\mu\rho_H}{\rho_H}\right)\\
\end{split}
\end{equation}
After transforming to $\{r,x^\mu\}$  (see appendix-\ref{app:horizon} for the details of the derivation)
\begin{equation}\label{eq:psir_0}
\begin{split}
\psi(r,x^\mu)
&=\left(\frac{r}{H}\right)^{1-\frac{1}{D}}+\left(\frac{r}{r_H}\right)^{1-\frac{1}{D}}\left(\frac{1}{r^2}-\frac{1}{r_H^2}\right)\frac{D-1}{D(D+1)}\left[\mathfrak{s}_1-\mathfrak{s}_2+\frac{1}{2}\left(\mathfrak{s}_3-\mathfrak{s}_4\right)+2~\mathfrak{s}_5\right]\\
&~~~~-\left(\frac{r}{r_H}\right)^{1-\frac{1}{D}}~\mathfrak{s}_4\left[\frac{2}{D(D-2)~r_H}\left(\frac{1}{r}-\frac{1}{r_H}\right)\right]+{\cal O}(\partial)^3\\
\end{split}
\end{equation}

\subsection{Fixing the normalization  of $\bar O^A$}
As we have explained in section-(\ref{sec:strategy}), the null geodesic field $\bar O^A\partial_A$ is related to the geodesic field $\bar O^A\partial_A $ (determined in section - (\ref{sec:hydsplit}) ) upto an overall normalization. The proportionality factor  $\Phi$ is given by the component of $\bar O_A$ in the direction of $n_A$-the unit normal to the constant $\psi$ hypersurfaces (see equation \eqref{eq:Phi}). More explicitly
\begin{equation}\label{eq:Phiexp}
\Phi \equiv\bar O^A n_A= n_r=\left(\partial_r\psi\over{\cal N}\right),~~\text{where}~~{\cal N}=\sqrt{(\partial_A\psi)~ \bar {\cal G}^{AB}~(\partial_B\psi)}
\end{equation}
However,  in $\{X^A\}$ coordinates it  is difficult to compute ${\cal N}$ and therefore $n_A$ since the background metric $\bar {\cal G}_{AB}$ and its inverse $\bar {\cal G}^{AB}$ are complicated.
Fortunately we also know $\psi$ in $\{Y^A\}$ coordinates where the background has a simple diagonal form. It is easier to compute $n_A$ first in $\{Y^A\}$ coordinates and then convert to $\{X^A\}$ coordinates.   Note in $\{X^A\}$ coordinates we only  need the $r$ component of $n_A$.\\ \\
In $\{Y^A\}$ coordinates :
\begin{equation}
n_{{A}} dY^{{A}}=\left[\frac{1}{\rho}-\frac{1}{2~\rho^3}\left(\frac{\partial_\mu \rho_H}{\rho_H}\right)^2\right]d\rho-\frac{1}{\rho_H(y)}\left(\frac{\partial \rho_H}{\partial y^\mu}\right)dy^\mu
\end{equation}\\
In $\{X^A\}$ coordinates :
\begin{equation}
\begin{split}
n_r&=\left(\frac{\partial \rho}{\partial r}\right)n_\rho+\left(\frac{\partial y^\mu}{\partial r}\right)n_\mu\\
&=\left[\frac{1}{\rho}-\frac{1}{2~\rho^3}\left(\frac{\partial_\mu \rho_H}{\rho_H}\right)^2\right]+\left[\frac{1}{r^2}u^\mu(x)+\frac{2}{r^3}\left(\frac{\Theta}{D-2}\right)u^\mu\right]\left[\frac{\partial_\mu \rho_H(y)}{\rho_H(y)}-\frac{1}{\rho}\left(\frac{\partial_\nu \rho_H}{\rho_H}\right)(\partial_\mu u^\nu)\right]
\end{split}
\end{equation}
After some simplifications the above expression becomes
\begin{equation}
n_r=\frac{1}{r}-\frac{1}{2~r^3}\left[\mathfrak{s}_1-\mathfrak{s}_2-4\left(\frac{r}{r_H}\right)\frac{\mathfrak{s}_4}{(D-1)(D-2)}+2\mathfrak{s}_5\right]=\Phi
\end{equation}
Substituting the normalization we get the following expression for $O_A$
\begin{equation}\label{eq:finO}
\begin{split}
O_A~dX^A =&~-r\left[1+\frac{1}{2r^2}\left(\mathfrak{s}_1-\mathfrak{s}_2-4\left(\frac{r}{r_H}\right)\frac{\mathfrak{s}_4}{(D-1)(D-2)}+2\mathfrak{s}_5\right)\right]u_\mu dx^\mu\\
\end{split}
\end{equation}

\subsubsection{Large-$D$ metric in terms of fluid data}
In section-(\ref{sec:largeDmetric}) we have described the large-$D$ metric upto corrections of order ${\cal O}\left(1\over D\right)^3$. It is written in terms of the extrinsic curvatures of the $\left(\psi=1\right)$ hypersurface and the derivatives of the membrane velocity field $U_A \equiv n_A -O_A$ .
Since $\psi$ and $O_A$ are already determined in terms of the fluid data, it is easy to express all the structures that appear in the large-$D$ metric in terms of the fluid data. We are listing it in  tables - (\ref{table:scalar_larged}), (\ref{table:tensor_larged}) and (\ref{table:vector_larged} ). 
\begin{table}[ht!]
\caption{Scalar large-$D$ Data in terms of fluid Data} % title of Table
\vspace{0.5cm}
\centering % used for centering table
\begin{tabular}{|c| c|} % centered columns (4 columns)
\hline\hline %inserts double horizontal lines
 Large-$D$ Data&Corresponding Fluid Data  \\ [1ex] % inserts table
%heading
\hline % inserts single horizontal line
\hline
\vspace{-0.3cm}
& \\
%\vspace{0.5cm}
${\cal S}_1\equiv\left(\frac{U\cdot\nabla K}{K}\right)$ & $=~0$\\ [1ex]
\hline
\vspace{-0.3cm}
& \\
${\cal S}_2\equiv U\cdot K\cdot U $&$=~-1+\left(\frac{1}{2r^2}\right)\left(\mathfrak{s}_1-\mathfrak{s}_2+2~\mathfrak{s}_5\right)$\\ [1ex]
\hline
\vspace{-0.3cm}
& \\
${\cal S}_3\equiv U\cdot K\cdot K\cdot U $ &$=~-1+\left(\frac{1}{r^2}\right)\left(\mathfrak{s}_1-\mathfrak{s}_2+2~\mathfrak{s}_5\right)$\\ [1ex]
\hline
\vspace{-0.3cm}
& \\
${\cal S}_4\equiv \Pi^{AB}\left(\frac{\nabla_A K}{K}\right)\left(\frac{\nabla_B K}{K}\right)$&$=~0$\\ [1ex]
\hline
\vspace{-0.3cm}
& \\
${\cal S}_5\equiv \Pi^{AB}\Pi^{CD}\left(\nabla_A U_C\right)\left(\nabla_B U_D\right)$&$=~\left(\frac{1}{r^2}\right)(\mathfrak{s}_4+\mathfrak{s}_3)$\\ [1ex]
\hline
\vspace{-0.3cm}
& \\
${\cal S}_6\equiv \Pi^{AB}[(U\cdot\nabla)U_A][(U\cdot\nabla)U_B]$&$=~0$\\ [1ex]
\hline
\vspace{-0.3cm}
& \\
${\cal S}_7\equiv U\cdot K\cdot\left(\frac{\nabla K}{K}\right)$&$=~0$\\ [1ex]
\hline
\vspace{-0.3cm}
& \\
${\cal S}_8\equiv \Pi^{AB}\left(\frac{\nabla_A K}{K}\right)[(U\cdot\nabla)U_B]$&$=~0$\\ [1ex]
\hline
\vspace{-0.3cm}
& \\
${\cal S}_9\equiv [(U\cdot\nabla)U^A][U^B K_{BA}]$&$=~0$\\ [1ex]
\hline
\vspace{-0.3cm}
& \\
${\cal S}_{10}\equiv \Pi^{AB}(\nabla_A U_B)$&$=~\frac{2}{r~r_H}\left(\frac{\mathfrak{s}_4}{D-1}\right)$\\ [1ex]
\hline
\vspace{-0.3cm}
& \\
${\cal S}_{11}\equiv \Pi^{AD}\Pi^{BC}\left(\nabla_A U_B\right)\left(\nabla_C U_D\right)$&$=~\frac{1}{r^2}(\mathfrak{s}_4-\mathfrak{s}_3)$\\ [1ex]
\hline
\hline
\end{tabular}
\label{table:scalar_larged} % is used to refer this table in the text
\end{table} 

\begin{table}[ht!]
\caption{Tensor large-$D$ Data in terms of fluid Data} % title of Table
\vspace{0.5cm}
\centering % used for centering table
\begin{tabular}{|c| c|} % centered columns (4 columns)
\hline\hline %inserts double horizontal lines
 Large-$D$ Data&Corresponding Fluid Data  \\ [1ex] % inserts table
%heading
\hline % inserts single horizontal line
\hline
\vspace{-0.3cm}
& \\
${\cal T}^{(1)}_{AB}~dX^A dX^B\equiv$& $=\bigg\{ r^2~ {\cal P}_{\mu\nu} + \left[ {\mathfrak s}_1-{\mathfrak s}_2 + 2{\mathfrak s}_5-\left(4\over(D-1)(D-2)\right)\left(r\over r_H\right){\mathfrak s}_4\right]{\cal P}_{\mu\nu} $\\
$P^C_AP^D_BP^{EF} \left(K_{EC} - \nabla_E U_C\right)$  & $ -2\left(r\over r_H\right)\left(D-2\over D-1\right)\left[u_\mu\left({\mathfrak v}_\nu^{(5)}-{\mathfrak v}^{(3)}_\nu\right)+u_\nu\left({\mathfrak v}_\mu^{(5)}-{\mathfrak v}^{(3)}_\mu\right)\right]-2~{\mathfrak t}^{(3)}_{\mu\nu}$\\
$\times\left(K_{FD} - \nabla_F U_D\right)~dX^A dX^B$&$+\left[u_\mu\left({\mathfrak v}_\nu^{(4)}+{\mathfrak v}^{(2)}_\nu-{\mathfrak v}^{(3)}_\nu\right)+u_\nu\left({\mathfrak v}_\mu^{(4)}+{\mathfrak v}^{(2)}_\mu-{\mathfrak v}^{(3)}_\mu\right)\right] \bigg\}dx^\mu dx^\nu$\\ [1ex]
\hline
\vspace{-0.3cm}
& \\
${\cal T}^{(2)}_{AB}~dX^A dX^B\equiv$&$=\bigg\{r^2 {\cal P}_{\mu\nu}+r\sigma_{\mu\nu}+{\cal P}_{\mu\nu}\left[\frac{\mathfrak{s}_1-\mathfrak{s}_2}{2}+\mathfrak{s}_5-\left(2\over(D-1)(D-2)\right)\left(r\over r_H\right){\mathfrak s}_4\right]$\\
$P^C_AP^D_B\left[K_{CD}-{\nabla_C U_D +\nabla_D U_C\over 2}\right]$&$+u_\mu\left[\mathfrak{v}_\nu^{(2)}-\mathfrak{v}_\nu^{(3)}+\mathfrak{v}_\nu^{(4)}+2\left(\frac{D-2}{D-1}\right)\frac{r}{r_H}\left(\mathfrak{v}_\nu^{(3)}-\mathfrak{v}_\nu^{(5)}\right)\right]$\\
$\times dX^A dX^B$&$+u_\nu\left[\mathfrak{v}_\mu^{(2)}-\mathfrak{v}_\mu^{(3)}+\mathfrak{v}_\mu^{(4)}+2\left(\frac{D-2}{D-1}\right)\frac{r}{r_H}\left(\mathfrak{v}_\mu^{(3)}-\mathfrak{v}_\mu^{(5)}\right)\right]\bigg\}dx^\mu dx^\nu$\\ [1ex]
\hline
\hline
\end{tabular}
\label{table:tensor_larged} % is used to refer this table in the text
\end{table}
\begin{table}[ht!]
\caption{Vector large-$D$ Data in terms of fluid Data} % title of Table
\vspace{0.5cm}
\centering % used for centering table
\begin{tabular}{|c| c|} % centered columns (4 columns)
\hline\hline %inserts double horizontal lines
 Large-$D$ Data&Corresponding Fluid Data  \\ [1ex] % inserts table
%heading
\hline % inserts single horizontal line
\hline
\vspace{-0.3cm}
& \\
%\vspace{0.5cm}
${\cal V}^{(1)}_A dX^A\equiv P^B_A\left(\frac{K^2}{D^2}\right)[(U\cdot\nabla)U_B]~dX^A$ & $=~\frac{(D-2)(D-1)}{D^2}\left(\frac{2}{r_H}\right)[\mathfrak{v}_\mu^{(5)}-\mathfrak{v}_\mu^{(3)}]dx^\mu$\\ [1ex]
\hline
\vspace{-0.3cm}
& \\
${\cal V}^{(2)}_A dX^A\equiv P^B_A\left(K^2\over D^2\right)\left(U^C K_{CB}\right)dX^A$&$=~\left(D-1\over D\right)^2\left(1\over r\right)\left[{\mathfrak v}_\mu^{(4)}+{\mathfrak v}_\mu^{(2)}  -{\mathfrak v}^{(3)}_\mu\right] dx^\mu$\\ [1ex]
\hline
\vspace{-0.3cm}
& \\
${\cal V}^{(3)}_A dX^A\equiv P^B_A P^F_D\left({\nabla_F K\over D}-{K\over D}U^EK_{EF}\right)\times$&$=-\left(D-1\over D\right)^2\left(1\over r\right)\left[{\mathfrak v}_\mu^{(4)}+{\mathfrak v}_\mu^{(2)}  -{\mathfrak v}^{(3)}_\mu\right] dx^\mu$ \\
~~~~~~~~~~~~~~~~~~$\left(K_{DB}-\nabla_D U_B\right)dX^A$ & \\[1ex]
\hline
\vspace{-0.3cm}
& \\
${\cal V}^{(4)}_A dX^A\equiv P^B_A\left(K^2\over D^2\right) \left(\nabla_B K\over K\right)dX^A$&$=~0$\\[1ex]
\hline
\vspace{-0.3cm}
& \\
${\cal V}^{(5)}_A dX^A\equiv P^A_C\left(\frac{\hat{\nabla}^2 U_A}{K}\right)dX^C$
&$=~\frac{1}{r}\bigg[2\left(\frac{D-2}{D-1}\right)\mathfrak{v}^{(5)}_\mu-\mathfrak{v}_\mu^{(3)}$\\
&~~~~~~~~~~~~~~$-\left(\frac{D-3}{D-1}\right)\left(\mathfrak{v}^{(2)}_\mu+\mathfrak{v}^{(4)}_\mu\right)\bigg]dx^\mu$\\[1ex]
\hline
\vspace{-0.3cm}
& \\
${\cal V}^{(6)}_A dX^A\equiv\frac{1}{{K}}P^A_C\left(U^B {{K}}_{B D} {{K}}^D_A\right)dX^C$
&$=~\frac{1}{r}\left(\frac{2}{D-1}\right)\left(\mathfrak{v}_\mu^{(4)}-\mathfrak{v}_\mu^{(3)}+\mathfrak{v}_\mu^{(2)}\right)dx^\mu$\\[1ex]
\hline
\vspace{-0.3cm}
& \\
${\cal V}^{(7)}_A dX^A\equiv\frac{1}{K}\Pi^{BA}\Pi^D_C\left(\frac{\nabla_B K}{K}\right)(\nabla_A U_D)dX^C$&$=~0$\\[1ex]
\hline
\vspace{-0.3cm}
& \\
${\cal V}^{(8)}_A dX^A\equiv P^A_C\left(\frac{\hat{\nabla}^2\hat{\nabla}^2 U_A}{{{K}}^3}\right)dX^C$
&$=~\left(\frac{1}{D-1}\right)^2 \frac{1}{r^5}\bigg[2\left(\frac{D-2}{D-1}\right)\mathfrak{v}^{(5)}_\mu-\mathfrak{v}_\mu^{(3)}$\\
&~~~~~~~~~~$-\left(\frac{D-3}{D-1}\right)\left(\mathfrak{v}^{(2)}_\mu+\mathfrak{v}^{(4)}_\mu\right)\bigg]dx^\mu$\\[1ex]
\hline
\vspace{-0.3cm}
& \\
${\cal V}^{(9)}_A dX^A\equiv\frac{1}{K^2}P^A_C K^{DE}\left(\hat{\nabla}_D\hat{\nabla}_E U_A\right)dX^C$
&$=~\left(\frac{1}{D-1}\right)\frac{1}{r}\bigg[2\left(\frac{D-2}{D-1}\right)\mathfrak{v}^{(5)}_\mu-\mathfrak{v}_\mu^{(3)}$\\
&~~~~~~~~$-\left(\frac{D-3}{D-1}\right)\left(\mathfrak{v}^{(2)}_\mu+\mathfrak{v}^{(4)}_\mu\right)\bigg]dx^\mu$\\[1ex]
\hline
\vspace{-0.3cm}
& \\
${\cal V}^{(10)}_A dX^A\equiv\frac{1}{K^3}P^A_C{\nabla}_A({{K}}_{BD} {{K}}^{BD} {{K}})dX^C$&$=~0$\\[1ex]
\hline
\vspace{-0.3cm}
& \\
${\cal V}^{(11)}_A dX^A\equiv P^A_C\left(\frac{\hat{\nabla}_A\hat{\nabla}^2{{K}}}{{{K}}^3}\right)dX^C$&$=~0$\\[1ex]
\hline
\hline
\end{tabular}
\label{table:vector_larged} % is used to refer this table in the text
\end{table}
Using these tables we can convert the scalar, vector and the tensor structures as described in equations \eqref{structurevecten} and \eqref{structurescalar} in terms of fluid data.
\begin{equation}\label{eq:convertscalar}
\begin{split}
S_1=~&\frac{1}{D}-\frac{1}{r^2}\left(2-\frac{1}{D}\right)(\mathfrak{s}_1-\mathfrak{s}_2+2\mathfrak{s}_5)-\frac{1}{r^2}\left(1-\frac{1}{D}\right)\mathfrak{s}_3-\frac{1}{r^2}\left(1+\frac{1}{D}\right)\mathfrak{s}_4\\
S_2=~&\frac{1}{D}\left(1-\frac{1}{D}\right)^2+\frac{2}{r^2}\left(\frac{(D-2)(D-1)}{D^3}\right)(\mathfrak{s}_1-\mathfrak{s}_2+2\mathfrak{s}_5)+\frac{1}{r^2}\left(\frac{(D-3)(D-1)}{D^3}\right)(\mathfrak{s}_3-\mathfrak{s}_4)\\
\end{split}
\end{equation}

\begin{equation}\label{eq:convertvec}
\begin{split}
V_r=~&0\\
V_\mu=~&-\frac{1}{r_H}\left(\frac{(D-2)(D-1)}{D^2}\right)(\mathfrak{v}_\mu^{(3)}-\mathfrak{v}_\mu^{(5)})\\
\end{split}
\end{equation}

\begin{equation}\label{eq:convertten}
\begin{split}
T_{rr}=~&0,~~~T_{r\mu}=~0\\
T_{\mu\nu}=~&-\left(\frac{r^2}{D}\right){\cal P}_{\mu\nu}+r\sigma_{\mu\nu}\left(1-\frac{1}{D}\right)+2~\mathfrak{t}^{(3)}_{\mu\nu}\\
&+{\cal P}_{\mu\nu}\left[-\frac{2}{D}(\mathfrak{s}_1-\mathfrak{s}_2+2\mathfrak{s}_5)-\frac{1}{D}(\mathfrak{s}_3-\mathfrak{s}_4)+\frac{r}{r_H}\left(1+\frac{1}{D}\right)\frac{2}{(D-2)(D-1)}\mathfrak{s}_4\right]\\
&+u_\mu\left[-\frac{1}{D}\left(\mathfrak{v}_\nu^{(2)}-\mathfrak{v}_\nu^{(3)}+\mathfrak{v}_\nu^{(4)}\right)-\frac{r}{r_H}\left(\frac{2(D-2)}{D(D-1)}\right)\left(\mathfrak{v}^{(3)}_\nu-\mathfrak{v}_\nu^{(5)}\right)\right]\\
&+u_\nu\left[-\frac{1}{D}\left(\mathfrak{v}_\mu^{(2)}-\mathfrak{v}_\mu^{(3)}+\mathfrak{v}_\mu^{(4)}\right)-\frac{r}{r_H}\left(\frac{2(D-2)}{D(D-1)}\right)\left(\mathfrak{v}^{(3)}_\mu-\mathfrak{v}_\mu^{(5)}\right)\right]
\end{split}
\end{equation}
Next we have to expand the functions (i.e., $f_1(R),~f_2(R),~v(R)$ and $t(R)$) appearing in the large-$D$ metric in terms of fluid data. Note that the arguments of these functions are $R\equiv D\left(\psi -1\right)$. Since $\psi$ admits an expansion in terms of derivatives so does these functions.\\
 Let us define a new variable $\tilde R \equiv D\left(\frac{r}{r_H} -1\right)$, which is of  zeroth order  derivative expansion. Now we express $R$ in terms of $\tilde R$.
 \begin{equation}\label{eq:tildeR}
 R =D\left[\left(1+\frac{\tilde R}{D}\right)^{1-\frac{1}{D}}-1\right] + \delta \tilde R
 \end{equation}
where,
\begin{equation}\label{eq:deltaR}
\begin{split}
\delta \tilde R&=-\left(1+\frac{\tilde R}{D}\right)^{-1-\frac{1}{D}}\frac{1}{r_H^2}\left(2~\tilde R+\frac{{\tilde R}^2}{D}\right)\frac{D-1}{D(D+1)}\left[\mathfrak{s}_1-\mathfrak{s}_2+2\mathfrak{s}_5+\frac{1}{2}(\mathfrak{s}_3-\mathfrak{s}_4)\right]\\
&~~~~-D\left(1-\frac{1}{D}\right)\left(1+\frac{\tilde R}{D}\right)^{1-\frac{1}{D}}\frac{1}{r_H^2}\bigg[h_1~\mathfrak{s}_4+h_2~\mathfrak{s}_3+(D-3)h_3(\mathfrak{s}_1-\mathfrak{s}_2+2~\mathfrak{s}_5)\bigg]\\
&~~~~+\left(1+\frac{\tilde R}{D}\right)^{-\frac{1}{D}}\frac{2~\tilde R}{D(D-2)}\frac{\mathfrak{s}_4}{r_H^2}
\end{split}
\end{equation}
Now the functions appearing in the large-$D$ metric could easily be expanded in derivative expansion upto the required order.
\begin{equation}\label{eq:expfn}
\begin{split}
&f_1(R) = f_1(\tilde{\bf R}) + \delta \tilde R ~\left(\partial f_1(\tilde{\bf R})\over\partial R\right),~~~
f_2(R) = f_2(\tilde{\bf R}) + \delta\tilde R~ \left(\partial f_2(\tilde{\bf R})\over\partial R\right)\\\
&v(R) = v(\tilde{\bf R}) + \delta\tilde R ~\left(\partial v(\tilde{\bf R})\over\partial R\right),~~~
t(R) = t(\tilde{\bf R}) + \delta\tilde R~ \left(\partial t(\tilde{\bf R})\over\partial R\right)\\\
&\text{where,} ~~\tilde{\bf R}=D\left[\left(1+\frac{\tilde R}{D}\right)^{1-\frac{1}{D}}-1\right] 
\end{split}
\end{equation}
In equation \eqref{eq:expfn} we did not explicitly evaluate the functions in terms of $\tilde R$ and we do not need to. Let us explain why.\\
We know the large-$D$ metric only  upto corrections of order ${\cal O}\left(1\over D\right)^3$. Also note that the functions $f_1(R)$, $f_2(R)$, $v(R)$ and $t(R)$ appear in the second order correction to the metric. In other words, whenever they occur, they always come with an explicit factor of $\left(1\over D\right)^2$. Therefore it follows that in equation \eqref{eq:expfn}, any term of the order ${\cal O}\left(1\over D\right)$ or higher is of no relevance. Expanding equation \eqref{eq:tildeR} further in $\left(1\over D\right)$  we find\footnote{When both $R$ and $\tilde R$ are of order $ {\cal O}\left(1\right)$  in terms of $\left(1\over D\right)$ expansion, in the functions we could simply replace $R$ by $\tilde R$. For regions,  where $R$ is of order $D$, we have to use the full relation as given as equation \eqref{eq:tildeR}. We could still neglect $\delta R$ but $R$ has to be replaced by $\tilde {\bf R}$ and not by $\tilde R$. However, as we have mentioned in a previous footnote, in these regions, the metric correction will fall exponentially with $D$ and therefore are not accurately captured by a power series expansion in $\left(1\over D\right)$.}
\begin{equation}\label{eq:Rexp}
R= \tilde {\bf R} +{1\over 2~ r_H^2}\left(1-\frac{\tilde {\bf R}}{D}\right)\left[{\mathfrak s}_3-{\mathfrak s}_4  + 2\left({\mathfrak s}_1 -{\mathfrak s}_2 + 2{\mathfrak s}_5\right)\right] + {\cal O}\left(1\over D\right)
\end{equation}
Here we have used the large-$D$ expansion of the coefficients $h_i$ appearing in equation \eqref{eq:deltaR}
\begin{equation}
\begin{split}
&h_1={1\over 2D}+ {\cal O}\left(1\over D\right)^2 ,~~~h_2=-\frac{1}{2D} + {\cal O}\left(1\over D\right)^2~~~\text{and,}~~~(D-3) h_3=-\frac{1}{D}+ {\cal O}\left(1\over D\right)^2\\
\end{split}
\end{equation}
Now examining the scalar, vector and the tensor structures in equations \eqref{eq:convertscalar}, \eqref{eq:convertvec}  and \eqref{eq:convertten}, we see that the terms are either of first or second order in terms of derivative expansion or of order ${\cal O}\left(1\over D\right) $ in terms of large-$D$ expansion. In either case the  ${\cal O}\left(\partial^2\right)$ terms in equation \eqref{eq:Rexp}, which are actually the leading terms of $\delta\tilde R$ in terms of $\left(1\over D\right)$ expansion, are negligible.\\
So, in $f_1(R)$, $f_2(R)$, $v(R)$ and $t(R)$ finally we could simply replace $R$ by $\tilde {\bf R}$.

Now we have all the ingredients to express the large-$D$ metric  particularly the `rest' part - ${\cal W}_{AB}^\text{rest}$ in terms of the fluid data. We substitute the data set presented  in  tables (\ref{table:scalar_larged}), (\ref{table:tensor_larged}) and  (\ref{table:vector_larged} )  in the metric described in section-(\ref{sec:largeDmetric}). By construction, ${\cal W}^\text{rest}_{rr}$ and ${\cal W}^\text{rest}_{r\mu}$ will vanish and only non-trivial components are ${\cal W}^\text{rest}_{\mu\nu}$. For convenience of comparison, we shall decompose the resultant expression for ${\cal W}_{\mu\nu}^\text{rest}$ again in  scalar, vector and the tensor sectors as we have done for ${\cal G}_{AB}^\text{rest}$.
\begin{equation}\label{eq:1byDfluidb}
\begin{split}
{\cal W}^\text{rest}_{\mu\nu}=~&{\cal W}_S^{(1)} u_\mu u_\nu + {\cal W}^{(2)}_S P_{\mu\nu} + \left({\cal W}^{(V)}_\mu u_\nu +{\cal W}^{(V)}_\nu u_\mu\right) + {\cal W}^{(T)}_{\mu\nu}
\end{split}
\end{equation}
where,
\begin{equation}\label{eq:grestpart_0}
\begin{split}
{\cal W}_S^{(1)}=~&r^2\left(\frac{r_H}{r}\right)^{D-1}-2\left(\mathfrak{s}_1-\mathfrak{s}_2+2\mathfrak{s}_5\right)\bigg[\frac{f_1(\tilde { R})}{D^2}+\frac{1}{D+1}\left(\frac{r_H}{r}\right)^{D-3}\left\{1-\left(\frac{r_H}{r}\right)^2\right\}\bigg]\\
&+\frac{\mathfrak{s}_3}{2}\left[-\left(\frac{r_H}{r}\right)^{D-3}-2~\frac{f_1(\tilde { R})}{D^2}+\left(\frac{r_H}{r}\right)^{D-3}\left(\frac{D-1}{D+1}\right)\left\{1-\left(\frac{r_H}{r}\right)^2\right\}\right]\\
&+\mathfrak{s}_4\left(\frac{r_H}{r}\right)^{D-3}\bigg[\frac{4}{(D-2)(D-1)}\left(1-\frac{r_H}{r}\right)-\left(\frac{r_H}{r}\right)^{-(D-3)}\left(\frac{f_1(\tilde{ R})}{D^2}\right)\\
&~~~~~~~~~~~~-\frac{2}{D-2}\left(1-\frac{r_H}{r}\right)-\frac{1}{2}\left(\frac{D-1}{D+1}\right)\left\{1-\left(\frac{r_H}{r}\right)^2\right\}-\frac{K_{2H}}{D-2}\bigg]\\
%{\cal W}_S^{(1)}=~&r^2\left(\frac{H}{r}\right)^{D-1}+\left(\frac{r_H}{r}\right)^{D-1}\left(\mathfrak{s}_1-\mathfrak{s}_2+2\mathfrak{s}_5-\frac{r}{r_H}\frac{4}{(D-2)(D-1)}~\mathfrak{s}_4\right)+2~\frac{f_1(R)}{D^2}\left[\mathfrak{s}_1-\mathfrak{s}_2+2\mathfrak{s}_5+\frac{1}{2}\left(\mathfrak{s}_3+\mathfrak{s}_4\right)\right]\\
%&+\left(\frac{r_H}{r}\right)^{D-2}\left[\left(1-\frac{r}{r_H}\right)\frac{2}{D-2}\mathfrak{s}_4-r~r_H\left(\frac{D-1}{D+1}\right)\left(\frac{1}{r^2}-\frac{1}{r_H^2}\right)\left(\mathfrak{s}_1-\mathfrak{s}_2+2\mathfrak{s}_5+\frac{1}{2}(\mathfrak{s}_3-\mathfrak{s}_4)\right)\right]+{\cal O}\left(\frac{1}{D}\right)^3\\
{\cal W}_S^{(2)}=~&{\cal O}\left(\frac{1}{D}\right)^3\\
{\cal W}^{(V)}_\mu=~&\frac{1}{D^2}\left(\frac{r}{r_H}\right)v(\tilde{ R})\left(\mathfrak{v}_\mu^{(3)}-\mathfrak{v}_\mu^{(5)}\right)+{\cal O}\left(\frac{1}{D}\right)^3\\
{\cal W}^{(T)}_{\mu\nu}=~&\frac{r~t(\tilde{ R})}{D^2}\sigma_{\mu\nu}+\left(\frac{2~t(\tilde{ R})}{D^2}\right)t^{(3)}_{\mu\nu}+{\cal O}\left(\frac{1}{D}\right)^3\\
\end{split}
\end{equation}
where, $~\tilde R \equiv D\left(\frac{r}{r_H} -1\right) $ and 
\begin{equation}\label{funcnrep}
\begin{split}
&t(R)=-~2\left(\frac{D}{K}\right)^2\int_R^{\infty}\frac{y~dy}{e^y-1}\\
&v(R)=2\left(\frac{D}{K}\right)^3\bigg[\int_R^{\infty}e^{-x}dx\int_0^x\frac{y~e^y}{e^y-1}dy~-~e^{-R}\int_0^{\infty}e^{-x}dx\int_0^x\frac{y~e^y}{e^y-1}dy\bigg]\\
&f_1(R)=-2\left(\frac{D}{K}\right)^2\int_R^{\infty}x~e^{-x}dx+2~e^{-R}\left(\frac{D}{K}\right)^2\int_0^{\infty}x~e^{-x}dx\\
\end{split}
\end{equation}
\begin{equation}
\begin{split}
&f_2(R)=\left(\frac{D}{K}\right)\Bigg[\int_R^{\infty}e^{-x}dx\int_0^x\frac{v(y)}{1-e^{-y}}dy-e^{-R}\int_0^{\infty}e^{-x}dx\int_0^x\frac{v(y)}{1-e^{-y}}dy\Bigg]\\
&~~~~~~~~-\left(\frac{D}{K}\right)^4\Bigg[\int_R^{\infty}e^{-x} dx\int_0^x\frac{y^2~ e^{-y}}{1-e^{-y}}dy-e^{-R}\int_0^{\infty}e^{-x} dx\int_0^x\frac{y^2 ~e^{-y}}{1-e^{-y}}dy\Bigg]\\
\end{split}
\end{equation}

\subsection{Comparison between ${\cal G}_{\mu\nu}^\text{rest}$ and ${\cal W}_{\mu\nu}^\text{rest}$}
We expect each component of ${\cal G}_{\mu\nu}^\text{rest}$ to be equal to ${\cal W}_{\mu\nu}^\text{rest}$ upto corrections of order ${\cal O}\left(\partial^3,~(1/D)^3\right)$. This would be true provided the coefficients ( functions of $r$ only) of independent scalar vector and tensor types of fluid data, appearing in both the metrics agree upto corrections of order ${\cal O}\left(1\over D\right)^3$. Below we are simply listing the equations that must be true for the  equality of the two metrics to be valid. In the next subsection we shall explicitly verify them by doing the integrations in the limit of large $D$.
\begin{table}[ht!]
\caption{Matching of ${\cal G}_{\mu\nu}^\text{rest}$ and ${\cal W}_{\mu\nu}^\text{rest}$} % title of Table
\vspace{0.5cm}
\centering % used for centering table
\begin{tabular}{|c| c|} % centered columns (4 columns)
\hline\hline %inserts double horizontal lines
Coefficient of different structures&The resultant equation\\ [1ex] % inserts table
%heading
\hline % inserts single horizontal line
\hline
\vspace{-0.3cm}
& \\
Coefficient of~~ $\sigma_{\mu\nu}$&$F({\bf r})=1+\frac{t(\tilde {R})}{2~D^2} + {\cal O}\left(1\over D\right)^3$\\[1ex]
\hline
\vspace{-0.3cm}
& \\
Coefficient of~~ $t^{(1)}_{\mu\nu}-{\cal P}_{\mu\nu}\left(\frac{\mathfrak{s}_4}{D-2}\right)$&$H_1({\bf r})=2~[F({\bf r})]^2-1=1+2~\frac{t(\tilde { R})}{D^2}+ {\cal O}\left(1\over D\right)^3$\\[1ex]
\hline
\vspace{-0.3cm}
& \\
Coefficient of~~ $t^{(3)}_{\mu\nu}$&$H_2({\bf r})=1+2~\frac{t(\tilde { R})}{D^2}+ {\cal O}\left(1\over D\right)^3$\\[1ex]
\hline
\vspace{-0.3cm}
& \\
Coefficient of~~ $t^{(4)}_{\mu\nu}+\left(\frac{\Theta}{D-2}\right)\sigma_{\mu\nu}$&$H_2({\bf r})=H_1({\bf r})+ {\cal O}\left(1\over D\right)^3$\\[1ex]
\hline
\vspace{-0.3cm}
& \\
Coefficient of~~ ${\cal P}_{\mu\nu}$&$K_1({\bf r})=2~[F({\bf r})]^2-1=1+2~\frac{t(\tilde{ R})}{D^2} + {\cal O}\left(1\over D\right)^3$\\[1ex]
\hline
\vspace{-0.3cm}
& \\
Coefficient of~~ $\mathfrak{v}_\mu^{(3)}-\mathfrak{v}_\mu^{(5)}$&$L({\bf r})=-\frac{1}{2D^3}\left(\frac{r}{r_H}\right){\bf r}^{D-3}v({\tilde { R}})$\\
&~~~~~~~~~~$=-\left(\frac{e^{\tilde R}}{2 D^3}\right)v(\tilde { R})+ {\cal O}\left(1\over D\right)^3$\\[1ex]
\hline
\vspace{-0.3cm}
& \\
Coefficient of~~ $\mathfrak{s}_1-\mathfrak{s}_2+2~\mathfrak{s}_5+\frac{\mathfrak{s}_3}{2}$&$f_1(\tilde R)=-2\tilde{R}~e^{-\tilde R}$\\[1ex]
\hline
\vspace{-0.3cm}
& \\
Coefficient of~~ $\mathfrak{s}_4$&$K_2({\bf r})=K_{2H}+\tilde R-\frac{1}{D}\left(4~\tilde R+\frac{3}{2}\tilde R^2\right)+ {\cal O}\left(1\over D\right)^3$\\[1ex]
\hline
\hline
\end{tabular}
\label{table:match} % is used to refer this table in the text
\end{table}

$$\text{Where}~~~{\bf r}\equiv{r\over r_H},~~~\tilde R \equiv D\left({\bf r} -1\right)$$

\subsubsection{$(1/D)$ expansion of the functions appearing in Hydrodynamic metric }\label{sec:largeDlimit}
In this subsection we shall verify the relations appearing in table (\ref{table:match}) upto order ${\cal O}\left(1\over D\right)^3$. For this, we need to evaluate the different integrals appearing in the hydrodynamic metric and expand it upto the required order in inverse power of dimension. For convenience we are quoting the integrals here again.
\begin{equation}\label{eq:notation2rep}
\begin{split}
&H_1(y) = 2y^2\int_y^\infty {dx\over x}\left[x^{D-3} -1\over x^{D-1} -1\right]\\
&H_2(y) = {F(y)^2}- 2~y^2\int _y ^\infty{dx\over x(x^{D-1}-1)}\int_1^x{dz\over z}\left[z^{D-3}-1\over z^{D-1}-1\right]\\
&K_1(y) = 2y^2\int_y^\infty {dx\over x^2}\int _x^\infty {dz\over z^2}\bigg[ z~ F'(z) - F(z)\bigg]^2\\
\end{split}
\end{equation}
\begin{equation}\nonumber
\begin{split}
&K_2(y)= \int_y^\infty\left(dx\over x^2\right)\bigg[1-2(D-2)~x^{D-2} -\left(1-{1\over x}\right)\bigg(xF'(x) - F(x)\bigg)\\
&~~~~~~~~~~~~~~~~~~~~~~~~~~~~~+\bigg(2(D-2)x^{D-1} - (D-3)\bigg)\int_x^\infty {dz\over z^2}\bigg(zF'(z) - F(z)\bigg)^2\bigg]\\
&L(y)=\int_y^\infty dx~x^{D-2}\int_x^\infty {dz\over z^3}\left[z-1\over z^{D-1} -1\right]
\end{split}
\end{equation}
Note that the expansion in inverse power of $D$ would crucially depend on how we choose to scale the variable $y$ or  the coordinate $\bf r$ with $D$. This is what we expect and we want a detailed match in the regime where $\left({\bf r}-1\right)\sim{\cal O}\left( {1\over D}\right)$.\\
 Below we shall first report the results of the integration in this regime, i.e., in equation \eqref{eq:notation2rep} we shall substitute $y= 1 +{Y\over D}$ with $Y\sim{\cal O}(1)$ and then evaluate the integral in an expansion in inverse powers of $D$ (see appendix (\ref{app:largeD}) for the details of the computation).
 
\vspace{1cm}
\underline{\textbf{Large $D$ expansion of  different functions in the `membrane region':}}
\begin{equation}\label{largeDlim1}
\begin{split}
F(y) &= F\left(1 +{Y\over D}\right) =~ 1 - \left(1\over D\right)^2\sum_{m=1}^\infty \left(1 + m Y\over m^2 \right)e^{-mY} + {\cal O}\left(1\over D^3\right)\\
H_1(y)&=H_1\left(1+\frac{Y}{D}\right)=~1-\left( 2\over D\right)^2\sum_{m=1}^\infty\left(1 +m Y\over m^2\right)e^{-mY} + {\cal O}\left(1\over D\right)^3\\
K_1(y)&=K_1\left(1+ {Y\over D}\right) =~1 -\left(1\over D\right)^3\sum_{m=1}^\infty \left(4\over m^3\right)(2 +mY) e^{-mY}+ {\cal O}\left(1\over D\right)^4\\
K_2(y)&=K_2\left(1+\frac{Y}{D}\right)=~-\left(\frac{D}{2}\right)+(3+Y) -\left(1\over 2D\right)\big[Y(8+3Y)\big] + {\cal O}\left(1\over D\right)^2\\
L(y)&=L\left(1+\frac{Y}{D}\right) =~  {\cal O}\left(1\over D\right)^3\\
H_2(y) &= H_2\left(1+\frac{Y}{D}\right)\\
&=1-\frac{1}{D^2}\bigg(\frac{\pi^2}{3}\left(e^Y-1\right)-4~Y~\text{Log}\left[1-e^{-Y}\right]+\left(e^Y-1\right)\left(~\text{Log}\left[1-e^{-Y}\right]~\right)^2\\
&~~~~~~+2~\left(e^Y-1\right)\text{Log}\left[1-e^{-Y}\right]~\text{Log}\left[\frac{1}{1-e^Y}\right]+2\left(e^Y+1\right)\text{PolyLog}[2,~e^{-Y}]\\
&~~~~~~-2\left(e^Y-1\right)\text{PolyLog}\left[2,~\frac{e^Y}{e^Y-1}\right]\bigg)+{\cal O}\left(\frac{1}{D}\right)^3\\
\end{split}
\end{equation}
Once we use this expansion in the equations we derived in the previous subsection, they are just trivially satisfied thus proving the equivalence of the two metrics within the membrane region.

Next we  have performed these integrations outside the membrane-region. In this region  $\left({\bf r}-1\right)\sim{\cal O}\left( 1\right)$, so here we have substitute $y = 1+\zeta$ with $\zeta\sim{\cal O}(1)$. It turns out that in this regime of $y$ , all the above functions evaluate to one upto corrections exponentially falling in $D$, and therefore non-perturbative from the point of view of $\left(1\over D\right)$ expansion (see appendix (\ref{app:largeD}) for the details of the computation).  Substituting this fact in the hydrodynamic metric, we see that outside the membrane region ${\cal G}_{AB}^\text{rest}$ vanishes exponentially fast in $D$, exactly as we have in case of large-$D$ metric.

\section{Implementing part-3:\\ Equivalence of the constraint equations}\label{sec:equieq}
In the previous subsection we have seen that the hydrodynamic metric is exactly same as the large-$D$ metric once we have correctly identified the membrane data of the large-$D$ expansion with the fluid data. However, as we have mentioned before, the matching of the two  metrics is not enough to show that the two gravity solutions are identical, since the time-evolution of  both the large-$D$ data and the fluid data are constrained  by two sets different looking equations. In this subsection our goal is to show these two sets of equations are also equivalent. More precisely what we would like to show is that  whenever  the fluid data would satisfy the appropriate relativistic Navier-Stokes equation, the corresponding  `membrane data'  would satisfy membrane equation.\\

The evolution equation of a fluid dual to $D$ dimensional gravity in presence of cosmological constant, could be expressed as a conservation of a stress tensor $T^{\mu\nu}_\text{fluid}$, living on the $(D-1)$ dimensional flat space-time. Up to first order in derivative expansion, $T^{\mu\nu}_\text{fluid}$ has the following structure, once expressed in terms of the fluid velocity $u^\mu$ and the temperature scale $r_H$.
\begin{equation}\label{eq:fluidstress}
\begin{split}
&T^{\mu\nu}_\text{fluid}=r_H^{D-1} \left[\left(D-1\right)u^\mu u^\nu+\eta^{\mu\nu}-\left(\frac{2}{r_H}\right)\sigma^{\mu\nu}\right] + {\cal O}\left(\partial^2\right)\\
\text{Fluid equation}:~~&\partial_\mu T^{\mu\nu}_\text{fluid}=0
\end{split}
\end{equation}

The membrane equation could also be expressed as a conservation of some membrane  stress tensor, living on the $D-1$ dimensional membrane embedded in the empty AdS.  \\
Suppose $\{z^a,~~a= 0,1,2,\cdots,D-2\}$ denotes the $(D-1)$ induced coordinates on the membrane. In terms of the membrane data (i.e., the membrane velocity $U^a$ and the membrane-shape encoded in the extrinsic curvature tensor $K_{ab}$) the stress tensor $\hat T^{ab}$ and conservation equation would have the following structure
\begin{equation}\label{eq:membranestress}
\begin{split}
 &\hat T^{ab}=\left(\frac{\cal K}{2}\right)U^a U^b+\left(\frac{1}{2}\right){\cal K}^{ab}-\frac{1}{2}\left(\tilde{\nabla}^a U^b+\tilde{\nabla}^b U^a\right)-\frac{1}{\cal K}\left(U^a\tilde{\nabla}^2U^b+U^b\tilde{\nabla}^2U^a\right)\\
&~~~~~~~~~~~~~~~~~+\frac{1}{2}\left(U^a\frac{\hat{\nabla}^b{\cal K}}{\cal K}+U^b\frac{\hat{\nabla}^a {\cal K}}{\cal K}\right)-\frac{1}{2}\left(U\cdot{\cal K}\cdot U+\frac{\cal K}{D}\right)g^{ab}_{\text{(ind)}} + {\cal O}\left(1\over D\right)\\
&\text{where}~~{\cal K} \equiv g_{(\text{ind})}^{ab}K_{ab},~~\tilde\nabla\equiv\text{Covariant derivative w.r.t. $g^{(\text{ind})}_{ab}$}\\
&\text{Membrane equation}:~~\tilde\nabla_a \hat T^{ab}=~0
\end{split}
\end{equation}
Now, we shall process the membrane equation and after rewriting $U^a$ and $K_{ab}$ in terms of the fluid data and their derivatives, we shall try to express the membrane equation as fluid equation plus terms identically vanishing upto the required order.\\
As before, for computational convenience we shall work in the $\{Y^A\} = \{\rho,y^\mu\}$ coordinates. In these coordinates, the location of the membrane is given by $\left[\rho -\rho_H(y) =0\right]$. Let us choose $\{z^a\}$ to be $\{y^\mu\}$ themselves so that the form of the induced metric is simple. Also with this choice of coordinates along the membrane,  there is no distinction between $\{a,b\}$ and $\{\mu,\nu\}$ indices and now onwards we shall use only the $\{\mu,\nu\}$ ones.

Note that since we are neglecting all terms of third or higher order in derivative expansion, and  since both fluid and the membrane equation already have one overall derivative, we need to know the membrane stress tensor $\hat T^{\mu\nu}$ only upto first order in derivative expansion. In \cite{prevwork} the membrane velocity $U^\mu$ and the membrane shape have already worked out upto first order in derivative expansion. We shall simply take their results and compute the other relevant quantities.
 \begin{equation}\label{eq:UShape}
 \begin{split}
 &g^\text{ind}_{\mu\nu} = r_H^2~ \eta_{\mu\nu}+{\cal O}(\partial)^2\\
&U_\mu~dy^\mu=r_H(y)~u_\mu(y)dy^\mu+{\cal O}(\partial)^2\\
&{\cal K}_{\mu\nu}=r_H^2(y)~\eta_{\mu\nu} +{\cal O}(\partial)^2,~~
{\cal K}=(D-1)+{\cal O}(\partial)^2,\\
&\frac{1}{2}\left(\tilde{\nabla}_\mu U_\nu+\tilde{\nabla}_\nu U_\mu\right)=r_H~\sigma_{\mu\nu}+{\cal O}(\partial)^2,~~~\hat{\nabla}^2 U_\mu={\cal O}(\partial)^2
 \end{split}
 \end{equation}

 Substituting equation \eqref{eq:UShape} in equation \eqref{eq:membranestress} we find
 \begin{equation}\label{eq:membraneFluid}
 \begin{split}
 \hat T^{\mu\nu}=\frac{1}{2}\left(\frac{1}{r_H^2}\right)\left[\left(D-1\right)u^\mu u^\nu+\eta^{\mu\nu}-\left(\frac{2}{r_H}\right)\sigma^{\mu\nu}\right]+{\cal O}\left(\frac{1}{D},\partial^2\right)
 \end{split}
 \end{equation}
 The Christoffel symbols are given by
 \begin{equation}
\begin{split}
\Gamma^\delta_{\beta\alpha}
&=\left[\delta^\delta_\alpha\left(\frac{\partial_\beta r_H}{r_H}\right)+\delta^\delta_\beta\left(\frac{\partial_\alpha r_H}{r_H}\right)-\eta_{\beta\alpha}\left(\frac{\partial^\delta r_H}{r_H}\right)\right]+{\cal O}(\partial)^3
\end{split}
\end{equation}
Rewriting  the membrane equation in terms $\partial_\mu$ and the Christoffel symbols we find
 \begin{equation}\label{eq:covderi}
 \begin{split}
 \tilde{\nabla}_\mu \hat T^{\mu\nu}&=\partial_\mu T^{\mu\nu}_\text{membrane}+\Gamma^\mu_{\mu\alpha}\hat T^{\alpha\nu}+\Gamma^\nu_{\mu\alpha}T^{\mu\alpha}_\text{membrane}\\
&=\partial_\mu\hat T^{\mu\nu}+(D+1)\left(\frac{\partial_\alpha r_H}{r_H}\right)\hat T^{\alpha\nu}-\left(\frac{\partial^\nu r_H}{r_H}\right)\left(\eta_{\alpha\beta}\hat T^{\alpha\beta}\right)+{\cal O}(\partial)^3\\
&=\left(1\over r_H^{D+1}\right)\partial_\mu \bigg[r_H^{D+1} ~\hat T^{\mu\nu}\bigg]-\left(\frac{\partial^\nu r_H}{r_H}\right)\left(\eta_{\alpha\beta}\hat T^{\alpha\beta}\right)+{\cal O}\left(1,\partial^3\right)\\
 \end{split}
 \end{equation}
Now substituting equation \eqref{eq:membraneFluid} in equation \eqref{eq:covderi} we find (upto corrections of order ${\cal O}(\partial^3) $ in derivative expansion and ${\cal O}(1)$ in large $D$ expansion)
 \begin{equation}\label{eq:finmatcheq}
 \begin{split}
 0=~&\tilde {\nabla}_\mu \hat T^{\mu\nu}\\
=~&\left(1\over r_H^{D+1}\right)\partial_\mu \bigg[r_H^{D+1} ~\hat T^{\mu\nu}\bigg]-\left(\frac{\partial^\nu r_H}{r_H}\right)\left(\eta_{\alpha\beta}\hat T^{\alpha\beta}\right)\\
=~&{1\over 2}\left(1\over  r_H^{D+1}\right)\partial_\mu \bigg(r_H^{D-1} \left[\left(D-1\right)u^\mu u^\nu+\eta^{\mu\nu}-\left(\frac{2}{r_H}\right)\sigma^{\mu\nu}\right]\bigg)\\
=~&{1\over 2}\left(1\over  r_H^{D+1}\right)\partial_\mu T^{\mu\nu}_\text{fluid}
 \end{split}
 \end{equation}
 Equation \eqref{eq:finmatcheq} clearly proves that upto the order  ${\cal O}\left(1/D^2,\partial^3\right)$, the two sets of constraint equations are equivalent once the data of the solution are appropriately identified with each other.\\
 
 Unfortunately, we still do not have the expression for the membrane stress tensor to second subleading order, though we do know the final membrane equation at this order (see equation \eqref{eq:constraint2}). Using the  tables (\ref{table:scalar_larged}) and (\ref{table:vector_larged})  we could easily rewrite each term of the membrane equation in terms of independent fluid data. We have checked (with the help of Mathematica version-11) that the membrane equation  vanishes  upto second subleading order provided the fluid equation is satisfied upto order ${\cal O}\left(\partial^2\right)$.

\section{Discussion and future directions}\label{sec:conclude}
In this note, we have compared two dynamical `black-hole' type solutions of Einstein's equations in presence of negative cosmological constant. These two solutions were already known and were determined using two different perturbation techniques - one is the `derivative expansion' and the other is an expansion in inverse powers of dimensions. We have shown that in the regime of overlap of the two perturbation parameters, the metric of these two  apparently different spaces are exactly same, to the order the solutions are known on both sides.\\
Very briefly our procedure is  as follows.\\
We have taken the metric generated in derivative expansion (known upto second order) in arbitrary number of space-time dimension-$D$ and expanded it in $\left(1\over D\right)$ upto order $\left(1\over D\right)^2$ assuming $D$ to be very large. \\
Next we have taken the  metric generated in  $\left(1\over D\right)$ expansion (this is also known upto second order) and expanded it in terms of boundary derivatives upto second order. The final result is that these two metrics just agree with each other.

The key reasons of this exact match  are the following.\\
 Firstly both the perturbation techniques use the same space-time ( namely the space-time of a Schwarzschild black-brane) as the starting point and secondly given the starting point and therefore  the characterizing  data, both of the techniques generate the higher order corrections uniquely. Hence in the regime where both perturbations are applicable, there must  exist just one solution with a given starting point. We could determine this solution by first applying derivative expansion and then expanding the answer further in ${\cal O}\left(1\over D\right)$ or vice versa.

As stated above, the matching of the two metrics seems quite simple in principle. But in practice it is quite complicated because the two metrics look very different from each other. In particular, unlike the hydrodynamic metric, the `large-$D$' metric is always generated in a `split' form - as a sum of `background'  which is just a pure AdS metric and the `rest' which is nontrivial only within the `membrane region' (a region of thickness of order ${\cal O}\left(1\over D\right)$ around the horizon).
The matching of these two metrics would  imply that from hydrodynamic metric if we subtract off its decaying part  (i.e, the part that falls off like $r^{-D}$ in the large $D$ limit), the remaining would be a metric for a regular space-time and would also satisfy the Einstein's equations in presence of negative cosmological constant. It does not follow just  from the `derivative expansion' technique.  In this note we have shown that this `non-decaying part of the hydrodynamic metric is actually  a coordinate transformation of pure AdS and this we have done without using any `large-$D$' expansion.\\
This is  one of the main result of this note.\\

This work could be  extended to several directions.\\
We have matched the two metrics only within the membrane region. But it  is possible to compute the gravitational radiation, sourced by the effective membrane stress tensor and extended outside the membrane region till infinity \cite{radiation}. In \cite{Dandekar:2017aiv} the authors have determined the boundary stress tensor from this radiation part and matched with the dual fluid stress tensor of the hydrodynamic metric. Now since we know how to `split' the hydrodynamic metric,  we could also match the metric coefficients outside the membrane region that are exponentially  falling off with $D$ and therefore non-perturbative from the point of view of ${\cal O}\left(1\over D\right)$ expansion.

We have compared the metric in the regime where both  perturbation techniques are applicable and what we have shown is that the derivative and the $\left(1\over D \right)$ expansion commute in this regime. However we also know \cite{prevwork} that there exists a regime where derivative expansion could not be applied but we could still apply `large-$D$' expansion. This is an interesting regime to explore since here we would construct genuinely new dynamical black hole solutions that were not described previously by other perturbative techniques. It would be very interesting to isolate out this regime in the general `large-$D$' expansion technique.

\section*{Acknowledgment}
It is a great pleasure to thank Shiraz Minwalla for initiating discussions on this topic.   We would  like to thank Yogesh Dandekar, Suman Kundu for illuminating discussions. P.B. and A.D. would like to acknowledge the hospitality of IISER-TVM while this work was in progress. P.B., A.D. and M.P. would like to acknowledge the hospitality at `Kavli Asian Winter School 2019', towards the final stages of this work.\\
%We would like to thank $\cdots$ for reading through the initial draft and very useful comments.\\
 We  would  also like to acknowledge our debt to the people of India for their steady and generous support to research in the basic sciences.

\appendix

\section{Comparison  upto ${\cal O}\left(\partial^2,{1\over D}\right)$ following \cite{prevwork} exactly:}\label{app:followprev}
As we have mentioned in the introduction, the computation of this note is quite different in its approach from that of \cite{prevwork} and in fact a bit conjectural in few steps. We could see these simple general patterns, conjectured to be true to all orders (for example see equations \eqref{eq:finOa} and \eqref{eq:allordermap}) only after we have done some detailed calculations, keeping all allowed arbitraryness  to begin with at every stage and fixing them one by one exactly the way it has been done in \cite{prevwork}. In this appendix, we shall record this way of doing the calculation. Though it looks clumsy, the clear advantage  in this brute-force method is that it is bound to give the correct result at the end.\\

Assuming derivative expansion, the most general form of $\bar{O}^A\partial_A$ at second order in derivative expansion which is null with respect to the full hydrodynamic metric ${\cal G}_{AB}$   is the following
\begin{equation}
\begin{split}
\bar{O}^A\partial_A =~&\partial_r+\left( \sum_{i=1}^5 B_i(r)~ \mathfrak{s}_{(i)} \right)\partial_r  + \sum_{i=1}^5 B_{5+i}(r)~\mathfrak{v}_{(i)}^\mu\partial_\mu
\end{split}
\end{equation}
Now imposing the condition that $\bar{O}_A$ is affinely parametrized null geodesic with respect to ${\cal G}_{AB}$ i.e., $(\bar O^A\nabla_A) \bar O^B =0$, the form of $\bar{O}^A$ becomes
\begin{equation}
\begin{split}
\bar{O}^A\partial_A =~&\partial_r +\left( \sum_{i=1}^5 b_i~ \mathfrak{s}_{(i)} \right)\partial_r  + \sum_{i=1}^5 \left(b_{5+i}\over r^2\right)~\mathfrak{v}_{(i)}^\mu\partial_\mu
\end{split}
\end{equation}
Where $b_i$'s are arbitrary constants.\\

As before,  we shall start with the most general possible form of $f^A$~'s upto second order in derivative expansion, substituing the answer for zeoth and first order correction from \cite{prevwork}.
\begin{equation}
\begin{split}
\rho &=r-\frac{\Theta}{D-2}+\sum_{i=1}^{5}c_{i}(r)~\mathfrak{s}_{i} \\
y^{\mu}&=x^{\mu}+\frac{u^\mu(x)}{r}+\left(\frac{\Theta}{D-2}\right)\left(\frac{u^{\mu}}{r^{2}}\right)+K^{\mu}_{\text{(old)}} +\left(\sum_{i=1}^{5}\tilde{c}_{i}(r)~\mathfrak{s}_{i}\right)u^{\mu}+\left(\sum_{i=1}^{5}\tilde{v}_{i}(r)~\mathfrak{v}_{i}^{\mu}\right)
\end{split}
\end{equation}
Where, $K^\mu_{\text{(old)}}$ is defined as
\begin{equation}
K^{\mu}_{\text{(old)}}=k_1(r_H)\left(\frac{\Theta}{D-2}\right)u^\mu+k_2(r_H) ~a^\mu
\end{equation}
We shall determine $f^A$ 's by using 
\begin{equation}
{\cal E}_B\equiv\bar{O}^A\left({\cal G}_{AB}-\bar{\cal G}_{AB}\right)=0
\end{equation}
Here, ${\cal G}_{AB}$ is full hydrodynamic metric in $\{X\}$ and $\bar{\cal G}_{AB}$ is background metric in $\{X\}$. Different components of $\bar{\cal G}_{AB}$ are as follows
\begin{equation}\label{eq:backX}
\begin{split}
 \bar {\cal G}_{rr}&=\frac{2}{r^{2}}\left[\sum_{i=1}^{5}\mathfrak{s}_{i}\left( c'_{i}(r)+r^{2}\tilde{c}'_{i}(r)- \frac{2c_{i}}{r}\right) +\left({3\over r^2}\right)\mathfrak{s}_1 \right]\\
   \bar {\cal G}_{r\mu}&= -u_{\mu} + \left(\frac{3}{r^{2}}\right)\mathfrak{s}_1~ u _{\mu} + u_{\mu} \sum_{i=1}^{5}\left(r^{2}\tilde{c}'_{i}-\frac{2c_{i}}{r}\right)\mathfrak{s}_{i} + \sum_{i=1}^{5}\left(r^{2}\tilde{v}'_{i}\mathfrak{v}^{i}_{\mu}\right)- u_{\beta}\partial_{\mu}K^{\beta}_{\text{(old)}} \\
%  \bar {\cal G}_{\mu\nu}&=\rho^2\bigg[\eta_{\mu\nu}+\frac{1}{r}\left(\partial_{\mu}u_{\nu}+\partial_{\nu}u_{\mu}\right)+\left(\frac{1}{r^2(D-2)}\right)\left(u_{\mu}\partial_{\nu}\Theta+u_{\nu}\partial_{\mu}\Theta\right)+\left(\frac{\Theta}{r^2(D-2)}\right)\left(\partial_{\mu}u_{\nu}+\partial_{\nu}u_{\mu}\right)\\
%&~~~~~~~~+\left(\frac{\partial_{\mu}u_{\beta}}{r}\right)\left(\frac{\partial_{\nu}u^{\beta}}{r}\right)+\partial_{\mu}K_{\nu}^{old}+\partial_{\nu}K_{\mu}^{old}\bigg]\\
%\bar{\cal G}_{\mu\nu} 
%&=r^2 \eta_{\mu\nu}-2r\left(\frac{\Theta}{D-2}\right)\eta_{\mu\nu} +r(\partial_\mu u_\nu+\partial_\nu u_\mu)+\eta_{\mu\nu}\left[\sum_{i=1}^5 2r~c_i(r)~\mathfrak{s}_i+\left(\frac{\Theta}{D-2}\right)^2\right]\\
%&~~~~-\left(\frac{\Theta}{D-2}\right)(\partial_\mu u_\nu+\partial_\nu u_\mu)+\left[u_\nu\partial_\mu\left(\frac{\Theta}{D-2}\right)+u_\mu\partial_\nu\left(\frac{\Theta}{D-2}\right)\right]\\
%&~~~~+(\partial_\mu u^\alpha)(\partial_\nu u_\alpha)+r^2\left[\partial_{\mu}K_{\nu}^{\text{(old)}}+\partial_{\nu}K_{\mu}^{\text{(old)}} \right]\\
\bar{\cal G}_{\mu\nu} 
&=r^2 ({\cal P}_{\mu\nu}-u_\mu u_\nu)+2r\left(\frac{\Theta}{D-2}\right)u_\mu u_\nu+r\left[2~\sigma_{\mu\nu}-a_\mu u_\nu-a_\nu u_\mu\right]\\
&+\left({\cal P}_{\mu\nu}-u_\mu u_\nu\right)\left[\sum_{i=1}^5 2r~c_i(r)\mathfrak{s}_i\right]-u_\mu u_\nu\left[\mathfrak{s}_1-\mathfrak{s}_2+2\mathfrak{s}_5\right]+u_\mu\left[\mathfrak{v}_\nu^{(2)}-\mathfrak{v}_\nu^{(3)}+\mathfrak{v}_\nu^{(4)}\right]\\
&+u_\nu\left[\mathfrak{v}_\mu^{(2)}-\mathfrak{v}_\mu^{(3)}+\mathfrak{v}_\mu^{(4)}\right]+\left[\mathfrak{t}^{(1)}_{\mu\nu}-\mathfrak{t}^{(2)}_{\mu\nu}+\mathfrak{t}^{(3)}_{\mu\nu}\right]+r^2\left[\partial_{\mu}K_{\nu}^{\text{(old)}}+\partial_{\nu}K_{\mu}^{\text{(old)}} \right]
\end{split}
\end{equation}
Solving, ${\cal E}_r=0$ and $u^\mu{\cal E}_\mu=0$ we get the following solution
\begin{equation}
\begin{split}
&c_1(r)=r\left(r_H\frac{\partial k_1}{\partial r_H}\right)+p_1,~~~
\tilde{c}_1(r)=\frac{1}{r^3}-\frac{p_1}{r^2}-\frac{r_H}{r}\left(\frac{\partial k_1}{\partial  r_H}\right)+\tilde{p}_1,\\                                                 
&c_2(r)=-k_2r+p_2,~~~
\tilde{c}_2(r)=\frac{k_2}{r}-\frac{p_2}{r^2}+\tilde{p}_2,\\
&c_3(r)=p_3,~~~
\tilde{c}_3(r)=-\frac{p_3}{r^2}+\tilde{p}_3,\\
\end{split}
\end{equation}
\begin{equation}\nonumber
\begin{split}
&c_4(r)=p_4,~~~
\tilde{c}_4(r)=-\frac{p_4}{r^2}+\tilde{p}_4,\\
&c_5(r)=-k_1r+p_5,~~~
\tilde{c}_5(r)=\frac{k_1}{r}-\frac{p_5}{r^2}+\tilde{p}_5.
\end{split}
\end{equation}
Solving, ${\cal P}^{\mu\nu}{\cal E}_\nu=0$
\begin{equation}
\begin{split}
\tilde{v}_1(r)&=\frac{1}{r}\left(k_2-r_H\frac{\partial k_1}{\partial r_H}\right)+q_1,~~~\tilde{v}_2(r)=-\frac{k_2}{r}+q_2,~~~\tilde{v}_3(r)=\frac{k_2}{r}+q_3\\
\tilde{v}_4(r)&=\frac{k_1}{r}+q_4,~~~\tilde{v}_5(r)=q_5
\end{split}
\end{equation} \\
Substituting the solutions for $c_i$, $\tilde c_i$ and $\tilde v_i$ it is easy to figure out the form  of background metric $\bar{\cal G}_{AB}$ in $\{X\}$ coordinates.
Now, to get the ${\cal G}_{AB}^\text{rest}$, we have to subtract off $\bar{\cal G}_{AB}$ from the hydrodynamic metric presented in section-(\ref{sec:hydmet}). The ${\cal G}_{rr}^\text{rest}$ and ${\cal G}_{r\mu}^\text{rest}$ simply vanish.
\begin{equation}\label{eq:grestrmu}
{\cal G}^\text{(rest)}_{rr}=0,~~{\cal G}_{r\mu}^{\text{(rest)}}=0
\end{equation}
The structure  of ${\cal G}_{\mu\nu}^\text{rest}$ is a bit complicated. We first decompose it into the scalar vector and the tensor sectors.
\begin{equation}
\begin{split}
{\cal G}_{\mu\nu}^{\text{(rest)}}={\cal G}^{(1)}_S u_\mu u_\nu+{\cal G}^{(2)}_S P_{\mu\nu}+\left({\cal G}^{(V)}_\mu u_\nu+{\cal G}^{(V)}_\nu u_\mu\right)+{\cal G}^{(T)}_{\mu\nu}
\end{split}
\end{equation}
Where, ${\cal G}^{(1)}_S$, ${\cal G}^{(2)}_S$, ${\cal G}^{(V)}_\mu$ and ${\cal G}^{(T)}_{\mu\nu}$ have the following forms
\begin{equation}\label{eq:grestscal}
\begin{split}
{\cal G}^{(1)}_S
&=r^2\left(\frac{r_H}{r}\right)^{D-1}+r^2\left(\frac{r_H}{r}\right)^{D-1}\left[-\frac{1}{r_H^2}\left(\frac{1}{2}-\frac{\tilde R}{D}\right)\mathfrak{s}_{(3)}+\frac{1}{r_H^2}\left(\frac{1}{2}-\frac{\tilde R+2}{D}\right)\mathfrak{s}_{(4)}\right]\\
&~~~~+2~r\left[\left(r~r_H~\frac{\partial k_1}{\partial r_H}+p_1\right)~\mathfrak{s}_{(1)}-(k_2~r-p_2)\mathfrak{s}_{(2)}+p_3~\mathfrak{s}_{(3)}+p_4~\mathfrak{s}_{(4)}-(k_1~r-p_5)\mathfrak{s}_{(5)}\right]\\
&~~~~-2~r^2\left[r_H\left(\frac{\partial k_1}{\partial r_H}\right)\mathfrak{s}_{(1)}-k_1~\mathfrak{s}_{(5)}-k_2~\mathfrak{s}_{(2)}\right]\\
{\cal G}_S^{(2)}
&=-2~r\bigg[\left(r~r_H~\frac{\partial k_1}{\partial r_H}+k_1~r+k_2~r+p_1\right)~\mathfrak{s}_{(1)}-(k_2~r-p_2)\mathfrak{s}_{(2)}+p_3~\mathfrak{s}_{(3)}\\
&~~~~~~~~~~~~+p_4~\mathfrak{s}_{(4)}-(k_1~r-k_2~r-p_5)\mathfrak{s}_{(5)}\bigg]\\
{\cal G}^{(V)}_\mu&=r^2 \left[(k_1-k_2)+r_H~\frac{\partial}{\partial r_H}(k_1-k_2)\right]\mathfrak{v}_\mu^{(1)}-(k_1-k_2)~r^2~\mathfrak{v}_\mu^{(4)}-2~k_2~r^2\left(\mathfrak{v}_\mu^{(3)}-\mathfrak{v}_\mu^{(2)}\right)
\\
{\cal G}^{(T)}_{\mu\nu}
&=-2~k_2~r^2~\mathfrak{t}^{(1)}_{\mu\nu}-2~k_1~r^2\left(\frac{\Theta}{D-2}\right)\sigma_{\mu\nu}
\end{split}
\end{equation}\\
Now, in the large-$D$ side the metric is quite simple if we neglect terms of order ${\cal O}\left(1\over D\right)^2$.
\begin{equation}\label{eq:1byD}
\begin{split}
G_{AB} =&~ \bar G_{AB} +G^\text{rest}_{AB}\\
\text{where}~~G^\text{rest}_{AB}=&~\left(\psi^{-D}\over \Phi^2\right)\bar O_A\bar O_B
\end{split}
\end{equation}\\
Where,
\begin{equation}\label{eq:psir}
\begin{split}
\psi^{-D}
&=\left(\frac{r}{H}\right)^{-D+1}-\left(\frac{r}{r_H}\right)^{-D+1}\left(\frac{1}{r^2}-\frac{1}{r_H^2}\right)\frac{D-1}{(D+1)}\left[\mathfrak{s}_1-\mathfrak{s}_2+\frac{1}{2}\left(\mathfrak{s}_3-\mathfrak{s}_4\right)+2~\mathfrak{s}_5\right]\\
&~~~~+\left(\frac{r}{r_H}\right)^{-D+1}~\mathfrak{s}_4\left[\frac{2}{(D-2)~r_H}\left(\frac{1}{r}-\frac{1}{r_H}\right)\right]+{\cal O}(\partial)^3\\
\Phi^2&=\frac{1}{r^2}-\frac{1}{r^4}\left[\mathfrak{s}_1-\mathfrak{s}_2-4\left(\frac{r}{r_H}\right)\frac{\mathfrak{s}_4}{(D-1)(D-2)}+2\mathfrak{s}_5\right]+{\cal O}(\partial)^3\\
\bar O_A~dX^A&=-u_\mu~dx^\mu-\left(\sum_{i=1}^5 b_i~ \mathfrak{s}_{(i)}~u_\mu-\sum_{i=1}^5 b_{5+i}~\mathfrak{v}_\mu^{(i)}\right)dx^\mu\\
\end{split}
\end{equation}
After using equation \eqref{eq:psir} we could rewrite $G^\text{rest}_{AB}$  in terms of  scalar,  vector and tensor type fluid data. 
\begin{equation}\label{eq:1byDfluid}
\begin{split}
G^\text{rest}_{rr}=~&0,~~G^\text{rest}_{r\mu}=0\\
G^\text{rest}_{\mu\nu}=~&G_S^{(1)} u_\mu u_\nu + G^{(2)}_S {\cal P}_{\mu\nu} + \left(G^{(V)}_\mu u_\nu +G^{(V)}_\nu u_\mu\right) + G^{(T)}_{\mu\nu}
\end{split}
\end{equation}
where
\begin{equation}\label{eq:grestpart}
\begin{split}
G_S^{(1)}=~&r^2\left(\frac{r_H}{r}\right)^{D-1}+r^2\left(\frac{r_H}{r}\right)^{D-1}\left[\frac{1}{r_H^2}\left(\frac{\tilde R}{D}-\frac{1}{2}\right)\mathfrak{s}_3+\frac{1}{r_H^2}\left(\frac{1}{2}-\frac{\tilde R+2}{D}\right)\mathfrak{s}_4+2\sum_{i=1}^5 b_i~ \mathfrak{s}_{(i)}\right]\\
G_S^{(2)}=~&0\\
G^{(V)}_\mu=~&-r^2\left(\frac{r_H}{r}\right)^{D-1}\left(\sum_{i=1}^5 {b_{5+i}}~\mathfrak{v}_\mu^{(i)}\right)\\
G^{(T)}_{\mu\nu}=~&0\\
\end{split}
\end{equation} 
Now we demand that each component of $G_{AB}^\text{rest}$ should be exactly equal to  the same component of ${\cal G}_{AB}^\text{rest}$.
We shall start from the tensor sector.
 $G^{(T)}_{\mu\nu}$ vanishes and therefore we shall set ${\cal G}^{(T)}_{\mu\nu}$ to zero implying
\begin{equation}\label{eq:cmpgt}
\begin{split}
k_1 =0;~~k_2 =0
\end{split}
\end{equation} 
 Now once we set $k_1$ and $k_2$ to zero, ${\cal G}^{(V)}_\mu$ vanishes. Therefore  $G^{(V)}_\mu$ also must vanish and  that determines the constants $b_{5+i} $s.
\begin{equation}\label{eq:cmpgv}
\begin{split}
b_{5 +i} =0,~~~i=\{1,\cdots,5\}
\end{split}
\end{equation}
Next we come to the comparison of ${\cal G}_{S}^{(2)}$ and $G_{S}^{(2)}$. After setting $k_1$ and $k_2$ to zero, ${\cal G}^{(2)}_s$ turns out to be
\begin{equation}\label{eq:cmpgtr}
{\cal G}^{(2)}_s= -2r\sum_{i=1}^5 p_i~ {\mathfrak s}_{(i)}
\end{equation}
whereas from equation \eqref{eq:grestpart} we see $G^{(2)}_S$ vanish. These two would be equal if we set all the constants $p_i$ s to zero.
\begin{equation}\label{eq:cmpgtr2}
\begin{split}
p_i =0,~~~i=\{1,\cdots,5\}
\end{split}
\end{equation}
Substituting equation \eqref{eq:cmpgt} and \eqref{eq:cmpgtr} in equation \eqref{eq:grestscal}, and equating ${\cal G}_s^{(1)}$ with $G_S^{(1)}$ we find
\begin{equation}\label{eq:cmpgtr2o}
\begin{split}
b_i =0,~~~i=\{1,\cdots,5\}
\end{split}
\end{equation}
The final form of $O^A$ becomes
\begin{equation}
O^A\partial_A=\partial_r+{\cal O}\left(\partial^3\right)
\end{equation}
The final form of the mapping functions are
\begin{equation}
\begin{split}
\rho &=r-\frac{\Theta}{D-2}+{\cal O}(\partial)^3\\
y^{\mu}&=x^{\mu}+\frac{u^\mu(x)}{\left(r-\frac{\Theta(x)}{D-2}\right)} +\left(\sum_{i=1}^{5}\tilde{p}_{i}~\mathfrak{s}_{i}\right)u^{\mu}+\left(\sum_{i=1}^{5}\tilde{q}_{i}~\mathfrak{v}_{i}^{\mu}\right)+{\cal O}(\partial)^3
\end{split}
\end{equation}
Where, $\tilde{p}_i$ and $\tilde{q}_i$ are some arbitrary constants.\\
They clearly match with \eqref{eq:finOa} and \eqref{eq:allordermap} upto the required order\footnote{The last two set of terms in the expression of $y^\mu$ do not get fixed by the matching at order ${\cal O}(\partial)^2$. They are equivalent to the terms $K^\mu_{\text{old}}$, which were ${\cal O}(\partial)$ terms in the expression of $y^\mu$ and did not get fixed from  matching at ${\cal O}(\partial)$. But $K^\mu_{\text{old}}$ do get fixed from  matching at order ${\cal O}(\partial)^2$ and which turns out to be zero. We think that these undetermined terms in $y^\mu$ will get fixed from the matching at the next order (i.e., ${\cal O}(\partial)^3$). We can very easily see that the expression of $y^\mu$ is consistent with \eqref{eq:finOa} $u^\mu(\frac{\partial\xi^\mu}{\partial x^\nu})={\cal O}(\partial)^3$.}

\section{Large-$D$ limit of the functions appearing  in hydrodynamic metric}\label{app:largeD}
In this appendix we shall evaluate the integrations appearing in equation \eqref{eq:notation2rep}. We are interested in some expressions, which could be further expanded in inverse powers of dimension.
But unfortunately we have not been able to do this integrations exactly in arbitrary $D$ and therefore we had to use several tricks to get to the answers, required for our comparison.

Integration would be done separately for two cases. One is for those ranges of $y$ so that the metric remains within the `membrane region'. Here we have to be careful so that we could fix each factor upto the corrections of order ${\cal O}\left(1\over D\right)^3$. This is the regime where we expect a detailed match between the large $D$ and the hydrodynamic metric. \\
Next we perform these integration outside the membrane regime. Here it is enough for us to show the overall fall off behaviour of these integrations as a function of $D$.

\subsection{Within the membrane region}
\subsubsection{$F(y):$}
\begin{equation}\label{eq:relF}
\begin{split}
&F(y)=1+\sum_{m=1}^\infty\left[\frac{y^{-m(D-1)}}{m(D-1)+1}-\frac{y^{-m(D-1)+1}}{m(D-1)}\right]\\
\end{split}
\end{equation}
Expanding in $\frac{1}{D}$, after putting $y=1+\frac{Y}{D}$,
\begin{equation}\label{F_y}
\begin{split}
F(y) =F\left(1 +{Y\over D}\right)&= 1 - \left(1\over D\right)^2\sum_{m=1}^\infty \left(1 + m Y\over m^2 \right)e^{-mY} + {\cal O}\left(1\over D^3\right)\\
&=1+\frac{1}{D^2}\bigg(Y~\text{Log}\left[1-e^{-Y}\right]-\text{PolyLog}\left[2,~e^{-Y}\right]\bigg)+{\cal O}\left(\frac{1}{D}\right)^3
\end{split}
\end{equation}
\subsubsection{$H_1(y):$}
Expanding the integrand:
\begin{equation}\label{eq:exH1stepa}
\begin{split}
Integrand =~& {1\over x}\left[x^{D-3} -1\over x^{D-1} -1\right]\\
=~&{1\over x^3}\left[1 - x^{-(D-3)}\right] \left[ \sum_{m=0}^\infty x^{-m(D-1)}\right]\\
=~&{1\over x^3}\left[ 1 + \sum_{m=1}^\infty\left(x^{-m(D-1) }\right) -  \sum_{m=1}^\infty\left(x^{-m(D-1) +2}\right) \right]
\end{split}
\end{equation}
After integration 
\begin{equation}\label{eq:exH1stepb_oo}
\begin{split}
H_1(y) =~&2y^2 \int_y^\infty{dx\over x}\left[x^{D-3} -1\over x^{D-1} -1\right]\\
=~&2y^2\int_y^\infty{dx\over x^3}\left[ 1 + \sum_{m=1}^\infty\left(x^{-m(D-1) }\right) -  \sum_{m=1}^\infty\left(x^{-m(D-1) +2}\right) \right]\\
=~&1 +2\sum_{m=1}^\infty \left[y^{-m(D-1)}\over m(D-1) +2\right] - 2\sum_{m=1}^\infty \left[y^{-m(D-1)+2}\over m(D-1)\right]
\end{split}
\end{equation}
Large-$D$ expansion after substituting~~~~ $y = 1+{Y\over D}$
\begin{equation}\label{eq:exH1stepb}
H_1\left(1+\frac{Y}{D}\right)=1-\left( 2\over D\right)^2\sum_{m=1}^\infty\left(1 +m Y\over m^2\right)e^{-mY} + {\cal O}\left(1\over D\right)^3
\end{equation}

\subsubsection{$K_1(y):$}
\begin{equation}\label{eq:relF_o}
F(y)=1+\sum_{m=1}^\infty\left[\frac{y^{-m(D-1)}}{m(D-1)+1}-\frac{y^{-m(D-1)+1}}{m(D-1)}\right]
\end{equation}

\begin{equation}\label{eq:relF2}
y F'(y) - F(y) = -1+(y-1)\sum_{m=1}^\infty y^{-m(D-1) } =-1 + {y-1\over y^{D-1}-1}
\end{equation}

\begin{equation}
\left[y F'(y) - F(y)\over y\right]^2 ={1\over y^2} +\sum_{m=1}^\infty \left[m \left(1-y\over y\right)^2y^{-(m+1)(d-1)} -2 \left(y-1\over y^2\right)y^{-m(D-1)}\right]
\end{equation}\\

\begin{equation}\label{eq:zfpz}
\begin{split}
&~~~~\int_y^\infty \left({dz\over z^2}\right)\bigg(z~F'(z) - F(z)\bigg)^2\\
&=\frac{1}{y}+\sum_{m=1}^\infty \left[\frac{1}{y}\left(\frac{2}{m(D-1)+1}\right)-\frac{2}{m(D-1}\right]y^{-m(D-1)}\\
&~~~~+\sum_{m=1}^\infty\left[-\frac{2~m}{(D-1)(m+1)}+\frac{1}{y}\left(\frac{m}{m(D-1)+D}\right)+y\left(\frac{m}{(D-1)m+D-2}\right)\right]y^{-(D-1)(m+1)}
\end{split}
\end{equation}

\begin{equation}\label{eq:Final K1}
\begin{split}
&~~~~K_1(x)\\
&=2 x^2\int_x^\infty {dy\over y^2}\int_y^\infty \left({dz\over z^2}\right)\bigg(z~F'(z) - F(z)\bigg)^2\\
&=1+4\sum_{m=1}^\infty\left[\frac{1}{[(D-1)m+1][(D-1)m+2]}-\frac{x}{[(D-1)m+1][m(D-1)]}\right]x^{-m(D-1)}\\
&~~~~+2\sum_{m=1}^\infty m\bigg[\frac{1}{[(D-1)m+D][D(m+1)-(m-1)]}-\frac{2~x}{[(D-1)(m+1)][D(m+1)-m]}\\
&~~~~~~~~~~~~+\frac{x^2}{[(D-1)(m+1)][D(m+1)-(m+2)]}\bigg]x^{-(D-1)(m+1)}
\end{split}
\end{equation}
Substituting $x = 1+ {X\over D}$ and taking the large-$D$ limit
\begin{equation}
K_1\left(1+ {X\over D}\right) =1 -\left(1\over D\right)^3\sum_{m=1} \left(4\over m^3\right)(2 +m ~X) e^{-m~X}+ {\cal O}\left(1\over D\right)^4
\end{equation}
\subsubsection{$K_2(y):$}
\begin{equation}\label{eq:K2rep}
\begin{split}
&K_2(y)= \int_y^\infty\left(dx\over x^2\right)\bigg[1-2(D-2)~x^{D-2} -\left(1-{1\over x}\right)\bigg(xF'(x) - F(x)\bigg)\\
&~~~~~~~~~~~~~~~~~~~~~~~~~~~~~+\bigg(2(D-2)x^{D-1} - (D-3)\bigg)\int_x^\infty {dz\over z^2}\bigg(zF'(z) - F(z)\bigg)^2\bigg]\\
\end{split}
\end{equation}
Naively the integration of $K_2(y)$ seems to be diverging. But upon appropriate expansion the singular terms cancel among themselves. After substituting \eqref{eq:relF2} and \eqref{eq:zfpz} in \eqref{eq:K2rep} we can integrate the whole expression.\\
\begin{equation}
\begin{split}
&~~~K_2(y)\\
&=\frac{2}{y}-\frac{1}{y^2}\left(\frac{D-2}{2}\right)\\
&~~~+\sum_{m=1}^\infty\bigg[\left(\frac{D(m-1)-3 m+2}{[(d-1)m][D(m+1)-(m+2)]}\right)+\frac{1}{y}\left(\frac{2 \left[-(D-3) m^2+2 (D-2) m+D-3\right]}{[(D-1) m (m+1)] [(D-1) m+1]}\right)\\
&~~~~~~~~~~~~+\frac{1}{y^2}\left(\frac{D^2 \left(m^2-3 m-2\right)+D \left(-4 m^2+10 m+5\right)+3 (m-3) m}{[(D-1) m+1] [(D-1) m+2] [D( m+1)-m]}\right)\bigg]y^{-m(D-1)}\\
&~~~+\sum_{m=1}^\infty\bigg[\frac{1}{y}\left(\frac{2 (D-3) m}{[(D-1) (m+1)] [D (m+1)-m]}\right)-\left(\frac{(D-3) m}{[(D-1) (m+1)] [D( m+1)-(m+2)]}\right)\\
&~~~~~~~~~~~~-\frac{1}{y^2}\left(\frac{(D-3) m }{[D( m+1)-m] [D( m+1)-(m-1)]}\right)\bigg]y^{-(D-1)(m+1)}\\
&~~~+\sum_{m=1}^\infty\bigg[\frac{1}{y}\left(\frac{4 (d-2)}{[d (m-1)-m+3] [(d-1) m+1]}\right)-\frac{4 (d-2) }{[(d-1) m] [d (m-1)-m+2]}\bigg]y^{-D(m-1)+(m-2)}
\end{split}
\end{equation}\\
Now, after putting $x=1+\frac{X}{D}$ we get
\begin{equation}
K_2\left(1+\frac{X}{D}\right)=-\left(\frac{D}{2}\right)+(3+X) -\left(1\over 2D\right)\big[X~(~8+3X~)\big] + {\cal O}\left(1\over D\right)^2
\end{equation}

\subsubsection{$L(y)$:}
\begin{equation}\label{eq:ly}
\begin{split}
L(y)=~&\int_y^\infty dx~x^{D-2}\int_x^\infty {dz\over z^3}\left[z-1\over z^{D-1} -1\right]\\
=~&\int_y^\infty dx~x^{D-2}\int_x^\infty dz\sum_{m=1}^\infty\left[z^{-m(D-1) -2} -z^{-m(D-1) -3}\right] \\
=~&\int_y^\infty dx~x^{D-2} \sum_{m=1}^\infty\left[{x^{-m(D-1) -1} \over m(D-1)+1}-{x^{-m(D-1) -2}\over m(D-1)+2}\right]\\
=~&\int_y^\infty dx~\sum_{m=0}^\infty\left[{x^{-m(D-1) -2} \over (m+1)(D-1)+1}-{x^{-m(D-1) -3}\over (m+1)(D-1)+2}\right]\\
=~&\sum_{m=0}^\infty\left[{y^{-m(D-1) -1} \over \left[(m+1)(D-1)+1\right]\left[m(D-1)+1\right]}-{y^{-m(D-1) -2}\over \left[(m+1)(D-1)+2\right] \left[m(D-1)+2\right]}\right]
\end{split}
\end{equation}
Substituting $y= 1 +{Y\over D}$ we find
\begin{equation}\label{eq:Lexp}
L\left(1+\frac{Y}{D}\right) =  {\cal O}\left(1\over D\right)^3
\end{equation}

\subsubsection{$H_2(y)$:}
\begin{equation}\label{eq:h2y}
\begin{split}
&H_2(y) = F(y)^2-2~y^2\int _y ^\infty{dx\over x(x^{D-1}-1)}\int_1^x{dz\over z}\left[z^{D-3}-1\over z^{D-1}-1\right]\\
\end{split}
\end{equation}
We shall first process the integral
\begin{equation}\label{eq:h2integral}
\begin{split}
&\int _y ^\infty{dx\over x(x^{D-1}-1)}\int_1^x{dz\over z}\left[z^{D-3}-1\over z^{D-1}-1\right]\\
=~&\bigg(\int_1^\infty{dz\over z}\left[z^{D-3}-1\over z^{D-1}-1\right]  \bigg) \int _y ^\infty{dx\over x(x^{D-1}-1)}- \int _y ^\infty{dx\over x(x^{D-1}-1)}\int_x^\infty{dz\over z}\left[z^{D-3}-1\over z^{D-1}-1\right]\\
=~&-{\cal Q}(D)\frac{\text{Log}[1-y^{-(D-1)}]}{D-1}- \int _y ^\infty{dx\over x(x^{D-1}-1)}\int_x^\infty{dz\over z}\left[z^{D-3}-1\over z^{D-1}-1\right]\\
\end{split}
\end{equation}
In the third step we have used  the following identities.
\begin{equation}\label{eq:intidnt}
\begin{split}
&\int_1^\infty{dz\over z}\left[z^{D-3}-1\over z^{D-1}-1\right]  = -\left(\frac{1}{D-1}\right)\left[ \text{EulerGamma} +\text{PolyGamma}\left(0,{2\over D-1}\right)\right] \equiv{\cal Q}(D)\\
&\int {dx\over x(x^{D-1}-1)} = {\text{Log}[1-x^{-(D-1)}]\over D-1}
\end{split}
\end{equation}
Now the second term could be processed in an expansion.
\begin{equation}
\begin{split}
\int _y ^\infty{dx\over x(x^{D-1}-1)}\int_x^\infty{dz\over z}\left[z^{D-3}-1\over z^{D-1}-1\right]&\equiv \int _y ^\infty dx\left[{1\over x(x^{D-1}-1)}\right] S(x)\\
\text{Where,}~~~ S(x)&\equiv \int_x^\infty{dz\over z}\left[z^{D-3}-1\over z^{D-1}-1\right]
\end{split}
\end{equation}
Now, first we will do the indefinite integral then will take the proper limit
\begin{equation}
\begin{split}
\int dx\left[\frac{S(x)}{x(x^{D-1}-1)}\right]&=S(x)\int \frac{dx}{x(x^{D-1}-1)}-\int dx\left(\frac{dS}{dx}\int\frac{dx}{x(x^{D-1}-1)}\right)\\
&=S(x)\int \frac{dx}{x(x^{D-1}-1)}+\int dx\left(\frac{x^{D-3}-1}{x(x^{D-1}-1)}\int\frac{dx}{x(x^{D-1}-1)}\right)\\
&=S(x)\int \frac{dx}{x(x^{D-1}-1)}+\int dx \left(x^{D-3}-1\right)\left(\frac{dG}{dx}\right)G(x)\\
\end{split}
\end{equation}
Where we have defined $$G(x)=\int\frac{dx}{x(x^{D-1}-1)}$$
So, we are getting
\begin{equation}
\begin{split}
&~~~~\int dx\left[\frac{S(x)}{x(x^{D-1}-1)}\right]\\
&=S(x)\int \frac{dx}{x(x^{D-1}-1)}+\frac{1}{2}\int dx \left(x^{D-3}-1\right)\frac{d}{dx}[G(x)]^2\\
&=S(x)\int \frac{dx}{x(x^{D-1}-1)}+\frac{1}{2}\int dx~\frac{d}{dx}\bigg[\left(x^{D-3}-1\right)[G(x)]^2\bigg]-\frac{1}{2}\int dx~[G(x)]^2(D-3)x^{D-4}\\
\end{split}
\end{equation}
Now,
\begin{equation}
G(x)\equiv \int\frac{dx}{x(x^{D-1}-1)}=\frac{\text{Log}[1-x^{-(D-1)}]}{D-1}
\end{equation}

\begin{equation}\label{int_H2}
\begin{split}
&~~~~\int dx\left[\frac{S(x)}{x(x^{D-1}-1)}\right]\\
&=S(x)\int \frac{dx}{x(x^{D-1}-1)}+\frac{1}{2~(D-1)^2}\left(x^{D-3}-1\right)\left(\text{Log}\left[1-x^{-(D-1)}\right]\right)^2\\
&~~~~~~~~-\frac{1}{2}\frac{D-3}{(D-1)^2}\int dx~x^{D-4}\left(\text{Log}\left[1-x^{-(D-1)}\right]\right)^2\\
&=S(x)\left(\frac{\text{Log}[1-x^{-(D-1)}]}{D-1}\right)+\frac{1}{2~(D-1)^2}\left(x^{D-3}-1\right)\left(\text{Log}\left[1-x^{-(D-1)}\right]\right)^2\\
&~~~~~~~~-\frac{1}{2}\frac{D-3}{(D-1)^2}\int dx~x^{D-4}\left(\text{Log}\left[1-x^{-(D-1)}\right]\right)^2\\
\end{split}
\end{equation}
Restoring the limit, we get
\begin{equation}
\begin{split}
&~~~~\int _y ^\infty{dx\over x(x^{D-1}-1)}\int_x^\infty{dz\over z}\left[z^{D-3}-1\over z^{D-1}-1\right]\\
&=\underbrace{\Bigg[S(x)\left(\frac{\text{Log}[1-x^{-(D-1)}]}{D-1}\right)\Bigg]_y^\infty}_{\text{Term1}}\underbrace{-\frac{1}{2~(D-1)^2}\left(y^{D-3}-1\right)\left(\text{Log}\left[1-y^{-(D-1)}\right]\right)^2}_{\text{Term2}}\\
&~~~~\underbrace{-\frac{1}{2}\frac{D-3}{(D-1)^2}\int_y^\infty dx~x^{D-4}\left(\text{Log}\left[1-x^{-(D-1)}\right]\right)^2}_{\text{Term3}}
\end{split}
\end{equation}
First we will calculate `Term3'. We can expand the integrand in `Term3' as follows
\begin{equation}
x^{D-4}\left(\text{Log}\left[1-x^{-(D-1)}\right]\right)^2=\sum_{m=2}^{\infty}2~ x^{D-4}x^{-m(D-1)}\frac{\text{HarmonicNumber}[m-1]}{m}
\end{equation}
Now, we can integrate term by term
\begin{equation}
\begin{split}
&~~~~ -\frac{1}{2}\frac{D-3}{(D-1)^2}\int_y^\infty dx~2~ x^{D-4}x^{-m(D-1)}\frac{\text{HarmonicNumber}[m-1]}{m}\\
&=-\frac{D-3}{(D-1)^2}\left(\frac{y^{-m(D-1)+D-3}~}{D(m-1)-m+3}\right)\frac{\text{HarmonicNumber}[m-1]}{m}
\end{split}
\end{equation}
Putting, $y=1+\frac{Y}{D}$
\begin{equation}
\begin{split}
&~~~~-\frac{D-3}{(D-1)^2}\left(\frac{y^{-m(D-1)+D-3}~}{D(m-1)-m+3}\right)\frac{\text{HarmonicNumber}[m-1]}{m}\\
&=\frac{1}{D^2}\left(\frac{e^{Y(1-m)}}{1-m}\right)\frac{\text{HarmonicNumber}[m-1]}{m}+{\cal O}\left(\frac{1}{D}\right)^3
\end{split}
\end{equation}
Now, if we take the summation we get,
\begin{equation}\label{Term3}
\begin{split}
\text{Term3}&=\sum_{m=2}^{\infty}\frac{1}{D^2}\left(\frac{e^{Y(1-m)}}{1-m}\right)\frac{\text{HarmonicNumber}[m-1]}{m}+{\cal O}\left(\frac{1}{D}\right)^3\\
&=-\frac{1}{6~D^2}\bigg[6~e^Y\text{PolyLog}\left[2,e^{-Y}\right]\\
&+(e^Y-1)\left(\pi^2+6~\text{Log}[1-e^{-Y}]~\text{Log}\left[\frac{1}{1-e^Y}\right]-6~\text{PolyLog}\left[2,\frac{e^Y}{e^Y-1}\right]\right)\bigg]+{\cal O}\left(\frac{1}{D}\right)^3
\end{split}
\end{equation}
Now we will calculate `Term2'
\begin{equation}
\text{Term2}=-\frac{1}{2~(D-1)^2}\left(y^{D-3}-1\right)\left(\text{Log}\left[1-y^{-(D-1)}\right]\right)^2
\end{equation}
Putting, $y=1+\frac{Y}{D}$
\begin{equation}\label{Term2}
\text{Term2}=-\frac{1}{2~D^2}\left(e^Y-1\right)\left(~\text{Log}\left[1-e^{-Y}\right]~\right)^2
\end{equation}
Now we will calculate `Term1'
\begin{equation}
\begin{split}
S(x)\left[\frac{\text{Log}[1-x^{-(D-1)}]}{D-1}\right]&=\left[\frac{\text{Log}[1-x^{-(D-1)}]}{D-1}\right]\int_x^\infty\frac{dz}{z}\left(\frac{z^{D-3}-1}{z^{D-1}-1}\right)\\
&=\left[\frac{\text{Log}[1-x^{-(D-1)}]}{D-1}\right]\int_x^\infty dz\sum_{m=0}^{\infty}\bigg[z^{-m(D-1)-3}-z^{-(m+1)(D-1)-1}\bigg]\\
&=\left[\frac{\text{Log}[1-x^{-(D-1)}]}{D-1}\right]\sum_{m=0}^{\infty}\left[\frac{x^{-m(D-1)-2}}{(D-1) m+2}-\frac{x^{-(D-1) (m+1)}}{(D-1) (m+1)}\right]
\end{split}
\end{equation}
\begin{equation}
\begin{split}
\text{Term1}&=\Bigg[S(x)\left(\frac{\text{Log}[1-x^{-(D-1)}]}{D-1}\right)\Bigg]_y^\infty\\
&=-\left[\frac{\text{Log}[1-y^{-(D-1)}]}{D-1}\right]\sum_{m=0}^{\infty}\left[\frac{y^{-m(D-1)-2}}{(D-1) m+2}-\frac{y^{-(D-1) (m+1)}}{(D-1) (m+1)}\right]\\
&=-\left[\frac{\text{Log}[1-y^{-(D-1)}]}{D-1}\right]\left[\frac{y^{-2}}{2}-\frac{y^{-(D-1)}}{D-1}\right]\\
&~~~~-\left[\frac{\text{Log}[1-y^{-(D-1)}]}{D-1}\right]\sum_{m=1}^{\infty}\left[\frac{y^{-m(D-1)-2}}{(D-1) m+2}-\frac{y^{-(D-1) (m+1)}}{(D-1) (m+1)}\right]\\
&=-\left[\frac{\text{Log}[1-y^{-(D-1)}]}{D-1}\right]\left[\frac{y^{-2}}{2}-\frac{y^{-(D-1)}}{D-1}\right]-\left[\frac{\text{Log}[1-y^{-(D-1)}]}{D-1}\right]\left[\frac{y^{-(D+1)}}{D-1}\right]\times\\
&~~~~~~~~~\left(y^2+\text{LerchPhi}\left[y^{1-D},~1,~\frac{D+1}{D-1}\right]+y^{D+1}\text{Log}[1-y^{-(D-1)}]\right)
\end{split}
\end{equation}
Now putting $y=1+\frac{Y}{D}$
\begin{equation}\label{Term1}
\text{Term1}=-\frac{1}{2~D}\text{Log}\left[1-e^{-Y}\right]+\frac{Y(Y+2)+2(e^Y-1)(2~Y-1)\text{Log}[1-e^{-Y}]}{4~D^2(e^Y-1)}+{\cal O}\left(\frac{1}{D}\right)^3
\end{equation}
Now, the integration \eqref{eq:h2integral} becomes
\begin{equation}
\begin{split}
&\int _y ^\infty{dx\over x(x^{D-1}-1)}\int_1^x{dz\over z}\left[z^{D-3}-1\over z^{D-1}-1\right]\\
&=-{\cal Q}(D)\frac{\text{Log}[1-y^{-(D-1)}]}{D-1}-\left[\text{Term1}+\text{Term2}+\text{Term3}\right]
\end{split}
\end{equation}
Putting, $y=1+\frac{Y}{D}$ we get
\begin{equation}\label{Q_D}
-{\cal Q}(D)\frac{\text{Log}[1-y^{-(D-1)}]}{D-1}=-\frac{\text{Log}[1-e^{-Y}]}{2~D}+\frac{Y(Y+2)-2~(e^Y-1)\text{Log}[1-e^{-Y}]}{4~D^2~(e^Y-1)}+{\cal O}\left(\frac{1}{D}\right)^3
\end{equation}
Now, the expression for $H_2(y)$
\begin{equation}\label{eq:H2}
\begin{split}
H_2(y) = &F(y)^2-2y^2\left[-{\cal Q}(D)\frac{\text{Log}[1-y^{-(D-1)}]}{D-1}-\left[\text{Term1}+\text{Term2}+\text{Term3}\right]\right]
\end{split}
\end{equation}
Putting \eqref{F_y}, \eqref{Q_D}, \eqref{Term1}, \eqref{Term2} and \eqref{Term3} we get the final expression
\begin{equation}
\begin{split}
H_2\left(1+\frac{Y}{D}\right)&=1-\frac{1}{D^2}\bigg(\frac{\pi^2}{3}\left(e^Y-1\right)-4~Y~\text{Log}\left[1-e^{-Y}\right]+\left(e^Y-1\right)\left(~\text{Log}\left[1-e^{-Y}\right]~\right)^2\\
&~~~~~~+2~\left(e^Y-1\right)\text{Log}\left[1-e^{-Y}\right]~\text{Log}\left[\frac{1}{1-e^Y}\right]+2\left(e^Y+1\right)\text{PolyLog}[2,~e^{-Y}]\\
&~~~~~~-2\left(e^Y-1\right)\text{PolyLog}\left[2,~\frac{e^Y}{e^Y-1}\right]\bigg)+{\cal O}\left(\frac{1}{D}\right)^3
\end{split}
\end{equation}
\subsection{Outside membrane region}
Here we shall show that fluid metric vanishes to any orders in $\frac{1}{D}$ expansion outside the membrane region. To show this we will use the following equation
\begin{equation}\label{main}
(1+\zeta)^{-(\alpha~D-\beta)}=e^{-(\alpha~D-\beta)~\text{Log}[1+\zeta]}
\end{equation}
Now, if $\zeta$ is some ${\cal O}(1)$ number then the right hand side is non perturbative in $\frac{1}{D}$ expansion. Now to show how the fluid metric behaves outside horizon we need to calculate $F(y), K_1(y), H_1(y)$ and $H_2(y)$ as the terms containing $L(y)$ and $K_2(y)$ are already multiplied by $y^{-(D-3)}$, hence non perturbative in $\frac{1}{D}$ expansion.
\subsubsection{$F(y)$:}
\begin{equation}
F(y)=y\int_y^\infty \frac{dx}{x}\left[\frac{x^{D-2}-1}{x^{D-1}-1}\right]
\end{equation}
For, $y=1+\zeta$, where $\zeta$ is some ${\cal O}(1)$ number, we can write the above integration \eqref{eq:relF} as 
\begin{equation}
\begin{split}
&~~~~F(1+\zeta)\\
&=1+\sum_{m=1}^\infty \left[\left(\frac{1}{(D-1)m+1}\right)(1+\zeta)^{-(D-1)m}-\left(\frac{1}{(D-1)m}\right)(1+\zeta)^{-(D-1)m+1}\right]\\
&=1+\sum_{m=1}^\infty \left[\left(\frac{1}{(D-1)m+1}\right)e^{-(m~D-m)~\text{Log}[1+\zeta]}-\left(\frac{1}{(D-1)m}\right)e^{-(m~D-m-1)~\text{Log}[1+\zeta]}\right]\\
&=1+ \text{terms non-perturbative in $\frac{1}{D}$}
\end{split}
\end{equation}
In the last line we have used \eqref{main}.
\subsubsection{$K_1(y)$:}
\begin{equation}
K_1(y)=~2 y^2\int_y^\infty {dx\over x^2}\int_x^\infty \left({dz\over z^2}\right)\bigg(z~F'(z) - F(z)\bigg)^2
\end{equation}
From \eqref{eq:Final K1}, for $y=1+\zeta$, where $\zeta$ is some ${\cal O}(1)$ number
\begin{equation}
\begin{split}
&~~~~K_1(1+\zeta)\\
&=1+4\sum_{m=1}^\infty\left[\frac{1}{[(D-1)m+1][(D-1)m+2]}-\frac{(1+\zeta)}{[(D-1)m+1][m(D-1)]}\right](1+\zeta)^{-m(D-1)}\\
&~~~~+2\sum_{m=1}^\infty m\bigg[\frac{1}{[(D-1)m+D][D(m+1)-(m-1)]}-\frac{2~(1+\zeta)}{[(D-1)(m+1)][D(m+1)-m]}\\
&~~~~~~~~~~~~+\frac{(1+\zeta)^2}{[(D-1)(m+1)][D(m+1)-(m+2)]}\bigg](1+\zeta)^{-(D-1)(m+1)}\\
\end{split}
\end{equation}
Now, using \eqref{main}, we get
\begin{equation}
K_1(1+\zeta)=1+ \text{terms non-perturbative in $\frac{1}{D}$}
\end{equation}
\subsubsection{$H_1(y)$:}
\begin{equation}
H_1(y) = 2y^2\int_y^\infty {dx\over x}\left[x^{D-3} -1\over x^{D-1} -1\right]
\end{equation}
From \eqref{eq:exH1stepb}, for $y=1+\zeta$, where $\zeta$ is some ${\cal O}(1)$ number
\begin{equation}\label{eq:exH1stepb_0}
\begin{split}
H_1(1+\zeta)&=1 +2\sum_{m=1}^\infty \left[(1+\zeta)^{-m(D-1)}\over m(D-1) +2\right] - 2\sum_{m=1}^\infty \left[(1+\zeta)^{-m(D-1)+2}\over m(D-1)\right]\\
%&=1+2\sum_{m=1}^\infty \left[e^{-(m~D-m)~\text{Log}[1+\zeta]}\over m(D-1) +2\right] - 2\sum_{m=1}^\infty \left[e^{-(m~D-m-2)~\text{Log}[1+\zeta]}\over m(D-1)\right]\\
\end{split}
\end{equation}
Now, using \eqref{main} we can very easily see that
\begin{equation}
H_1(1+\zeta)=1+ \text{terms non-perturbative in $\frac{1}{D}$}
\end{equation}
\subsubsection{$H_2(y)$:}
\begin{equation}
\begin{split}
&H_2(y) = F(y)^2-2~y^2\int _y ^\infty{dx\over x(x^{D-1}-1)}\int_1^x{dz\over z}\left[z^{D-3}-1\over z^{D-1}-1\right]\\
\end{split}
\end{equation}
From\eqref{eq:H2}, for $y=1+\zeta$, where $\zeta$ is some ${\cal O}(1)$ number
\begin{equation}
\begin{split}
H_2(y) = &F(y)^2-2y^2~{\cal Q}(D)\left[\sum_{m=1}^{\infty}\frac{y^{-m(D-1)}}{m(D-1)}\right]+2~y^2\left[\text{Term1}+\text{Term2}+\text{Term3}\right]\\
\end{split}
\end{equation}
where,
\begin{equation}
\begin{split}
{\cal Q}(D) &= -\left(\frac{1}{D-1}\right)\left[ \text{EulerGamma} +\text{PolyGamma}\left(0,{2\over D-1}\right)\right]\\
\text{Term1}&=\left[\sum_{m=1}^\infty\frac{y^{-m(D-1)}}{m(D-1)}\right]\left[\frac{y^{-2}}{2}-\frac{y^{-(D-1)}}{D-1}\right]\\
&~~~~+\left[\sum_{m=1}^\infty\frac{y^{-m(D-1)}}{m(D-1)}\right]\left(\frac{y^{-(D-1)}}{D-1}+y^{-(D+1)}\sum_{n=0}^\infty\frac{y^{-n(D-1)}}{n(D-1)+(D+1)}-\sum_{n=1}^\infty\frac{y^{-n(D-1)}}{n(D-1)}\right)\\
\end{split}
\end{equation}
\begin{equation}
\begin{split}
\text{Term2}&=-\frac{1}{2~(D-1)^2}\left(y^{D-3}-1\right)\left(\sum_{m,n=1}^\infty\frac{y^{-(m+n)(D-1)}}{m ~n}\right)\\
\text{Term3}&=-\frac{D-3}{(D-1)^2}\sum_{m=2}^\infty\left(\frac{y^{-m(D-1)+D-3}~}{D(m-1)-m+3}\right)\frac{\text{HarmonicNumber}[m-1]}{m}
\end{split}
\end{equation}
Now using \eqref{main} we can easily see that
\begin{equation}
H_2(1+\zeta)=1+ \text{terms non-perturbative in $\frac{1}{D}$}
\end{equation}
Using the above results in \eqref{eq:grestmunu} we see that ${\cal G}_{\mu\nu}^{\text{rest}}$ vanishes for all order in $\frac{1}{D}$ expansion outside the membrane region.

\section{Relation between Horizon $\rho_H(y)$ in $Y^A(\equiv\{\rho,y^\mu\})$ coordinates and $H(x)$ in $X^A(\equiv\{r,x^\mu\} )$ coordinates:}\label{app:horizon}
In this appendix we shall determine the relation between the position of the horizon($\rho_H(y^\mu)$) in $Y^A$ - coordinates with the position of the horizon ($H(x^\mu)$) in $X^A$ - coordinates. ($X^A\equiv\{r,x^\mu\}$) and ($Y^A\equiv\{\rho,y^\mu\}$) are related through the following coordinate transformation
\begin{equation}
\rho = r - \left(\Theta(x)\over D-2\right),~~~y^\mu = x^\mu + {u^\mu(x) \over r -{\Theta(x)\over D-2}}
\end{equation}
The inverse of the above coordinate transformation is
\begin{equation}
\begin{split}
r&=\rho+\left(\frac{\Theta(y)}{D-2}\right)-\frac{1}{\rho}(u\cdot\partial)\left(\frac{\Theta}{D-2}\right)+{\cal O}(\partial)^3\\
x^\mu&=y^\mu-\frac{u^\mu(y)}{\rho}+\frac{a^\mu(y)}{\rho^2}+{\cal O}(\partial)^2
\end{split}
\end{equation}
Now, the equation of the horizon is 
\begin{equation}
\begin{split}
r&=H(x)\\
\Rightarrow ~ \rho&=H\left(y^\mu-\frac{u^\mu(y)}{\rho}+\frac{a^\mu(y)}{\rho^2}\right)-\left(\frac{\Theta(y)}{D-2}\right)+\frac{1}{\rho}(u\cdot\partial)\left(\frac{\Theta}{D-2}\right)+{\cal O}(\partial)^3\\
&=H(y)-\frac{(u\cdot\partial)H(y)}{\rho}+\frac{(a\cdot\partial)H(y)}{\rho^2}+\frac{u^\mu u^\nu\partial_\mu\partial_\nu H(y)}{2\rho^2}-\left(\frac{\Theta(y)}{D-2}\right)\\
&~~~~~~~~~~~~+\frac{1}{\rho}(u\cdot\partial)\left(\frac{\Theta}{D-2}\right)+{\cal O}(\partial)^3\\
\end{split}
\end{equation}
Using \eqref{eq:hor} we can write the equation of the horizon as
\begin{equation}\label{eqn1}
\begin{split}
\rho&=H(y)-\left(\frac{\Theta(y)}{D-2}\right)-\frac{(u\cdot\partial)r_H(y)}{\rho}+\frac{(a\cdot\partial)r_H(y)}{\rho^2}+\frac{u^\mu u^\nu\partial_\mu\partial_\nu r_H(y)}{2\rho^2}\\
&~~~~~~~~~~~~+\frac{1}{\rho}(u\cdot\partial)\left(\frac{\Theta}{D-2}\right)+{\cal O}(\partial)^3\\
\end{split}
\end{equation}
Up to ${\cal O}(\partial)^2$ the equation of the horizon is
\begin{equation}\label{eqn3}
\begin{split}
\rho&=H(y)-\left(\frac{\Theta(y)}{D-2}\right)-\frac{(u\cdot\partial)r_H(y)}{r_H(y)}+{\cal O}(\partial)^2\\
&=r_H(y)+{\cal O}(\partial)^2
\end{split}
\end{equation}
Where, In the last line we have used \eqref{NSscalar} and \eqref{eq:hor}.\\
Using, \eqref{eqn3} in \eqref{eqn1} we get,
\begin{equation}
\begin{split}
\rho&=H(y)-\left(\frac{\Theta(y)}{D-2}\right)-\frac{(u\cdot\partial)r_H(y)}{r_H(y)}+\frac{(a\cdot\partial)r_H(y)}{r^2_H(y)}+\frac{u^\mu u^\nu\partial_\mu\partial_\nu r_H(y)}{2~r^2_H(y)}\\
&~~~~~~~~~~~~+\frac{1}{r_H(y)}(u\cdot\partial)\left(\frac{\Theta}{D-2}\right)+{\cal O}(\partial)^3\\
\end{split}
\end{equation}
After some simplifications the above expression becomes
\begin{equation}\label{eqn4}
\rho=H(y)+\frac{1}{2~r_H}(u\cdot\partial)\left(\frac{\Theta}{D-2}\right)+\frac{1}{2~r_H}\left(\frac{\Theta}{D-2}\right)^2-\frac{a^2}{2~r_H}-\frac{2}{r_H}\frac{\sigma^2}{(D-1)(D-2)}+{\cal O}(\partial)^3\\
\end{equation}
In ($Y^A\equiv\{\rho,y^\mu\}$) coordinate the equation of the horizon is
\begin{equation}\label{eqn2}
\rho=\rho_H(y)
\end{equation}
From \eqref{eqn4} and \eqref{eqn2} we get
\begin{equation}\label{eqn5}
\begin{split}
\rho_H(y)&=H(y)+\frac{1}{2~r_H}(u\cdot\partial)\left(\frac{\Theta}{D-2}\right)+\frac{1}{2~r_H}\left(\frac{\Theta}{D-2}\right)^2-\frac{a^2}{2~r_H}-\frac{2}{r_H}\frac{\sigma^2}{(D-1)(D-2)}+{\cal O}(\partial)^3
\end{split}
\end{equation}
We can express $H(y)$ in terms of $H(x)$
\begin{equation}\label{eqn6}
\begin{split}
H(y)&=H(x)-\frac{r_H(x)}{r}\left(\frac{\Theta(x)}{D-2}\right)-\frac{r_H}{r^2}\left(\frac{\Theta(x)}{D-2}\right)^2+\frac{2}{r}\left(\frac{\sigma^2}{(D-1)(D-2)}\right)\\
&~~~~+\frac{r_H}{2~r^2}\left[\left(\frac{\Theta}{D-2}\right)^2+a^2-(u\cdot\partial)\frac{\Theta}{D-2}\right]+{\cal O}(\partial)^3
\end{split}
\end{equation}
Substituting \eqref{eqn6} in \eqref{eqn5} we get the final expression
\begin{equation}\label{rhohx}
\begin{split}
 \rho_H(y)&=H(x)-\frac{r_H(x)}{r}\left(\frac{\Theta(x)}{D-2}\right)-\frac{2}{r_H}\left(1-\frac{r_H}{r}\right)\frac{\mathfrak{s}_4}{(D-1)(D-2)}\\
 &~~~~+\frac{1}{2~r_H}\left(1-\frac{r_H^2}{r^2}\right)\left[\mathfrak{s}_1-\mathfrak{s}_2+\mathfrak{s}_5\right]+{\cal O}\left(\partial\right)^3\\
\end{split}
\end{equation}

\section{Identities}\label{app:identities}
In this appendix we shall give the derivation of the identities we have used in subsection \ref{sec:largeDfluid}.

\subsection*{Identity 1:}
\begin{equation}\label{identity1}
\begin{split}
\left(\frac{\partial_\mu r_H}{r_H}\right)^2&=\left[-a_\mu+u_\mu\left(\frac{\Theta}{D-2}\right)\right]\left[-a^\mu+u^\mu\left(\frac{\Theta}{D-2}\right)\right]+{\cal O}\left(\partial\right)^3\\
&=a^2-\left(\frac{\Theta}{D-2}\right)^2+{\cal O}\left(\partial\right)^3\\
&=\mathfrak{s}_2-\mathfrak{s}_1+{\cal O}\left(\partial\right)^3
\end{split}
\end{equation}
\subsection*{Identity 2:}
\begin{equation}\label{identity2}
\begin{split}
\left(\frac{\partial^\mu\partial_\mu r_H}{r_H}\right)&=\frac{1}{r_H}\partial^\mu\left[-r_H~a_\mu+r_H~u_\mu\left(\frac{\Theta}{D-2}\right)\right]\\
&=-(\partial\cdot a)-\frac{(a\cdot\partial)r_H}{r_H}+\frac{\Theta^2}{D-2}+\left(\frac{\Theta}{D-2}\right)\left(\frac{(u\cdot\partial)r_H}{r_H}\right)+(u\cdot\partial)\left(\frac{\Theta}{D-2}\right)\\
&=-\left(\sigma^2-\omega^2+\frac{\Theta^2}{D-2}+(u\cdot\partial)\Theta\right)+a^2+\frac{\Theta^2}{D-2}-\left(\frac{\Theta}{D-2}\right)^2\\
&~~~~~~~~~~~~~~+(u\cdot\partial)\left(\frac{\Theta}{D-2}\right)+{\cal O}\left(\partial\right)^3\\
&=\omega^2-\sigma^2-(D-3)(u\cdot\partial)\left(\frac{\Theta}{D-2}\right)+a^2-\left(\frac{\Theta}{D-2}\right)^2+{\cal O}\left(\partial\right)^3\\
&=\mathfrak{s}_3-\mathfrak{s}_4-(D-3)\mathfrak{s}_5+\mathfrak{s}_2-\mathfrak{s}_1+{\cal O}\left(\partial\right)^3
\end{split}
\end{equation}
\subsection*{Identity 3:}
In this identity we shall just quote the fluid equation upto second subleading order. The derivation is quite straight forward. We have to calculate divergence of fluid stress tensor \eqref{eq:fluidstress_1} and have to project it along $u_\mu$ direction  and perpendicular to $u^\mu$ direction. 
\begin{equation}\label{NSscalar}
\begin{split}
&u_\nu\partial_\mu(T^{\mu\nu})={\cal O}\left(\partial\right)^3\\
\Rightarrow~~&\frac{(u\cdot\partial)r_{\text{H}}}{r_{H}}=-\frac{\Theta}{D-2}+\left(\frac{2}{r_{\text{H}}}\right)\frac{\sigma^2}{(D-1)(D-2)}+{\cal O}\left(\partial\right)^3
\end{split}
\end{equation}
\begin{equation}\label{NSvector}
\begin{split}
&{\cal P}_\nu^\alpha\partial_\mu(T^{\mu\nu})={\cal O}\left(\partial\right)^3\\
\Rightarrow~~&{\cal P}^\alpha_\nu~\left(\frac{\partial^\nu r_H}{r_H}\right)={\cal P}^\alpha_\nu\left[-a^\nu-\frac{2}{r_H}\left(\frac{D-2}{D-1}\right)(a_\mu\sigma^{\mu\nu})+\frac{2}{r_H}\left(\frac{1}{D-1}\right)\partial_\mu \sigma^{\mu\nu}\right]+{\cal O}\left(\partial\right)^3
\end{split}
\end{equation}
\newpage
\section{Notations}\label{app:notation}
\begin{table}[ht]
\caption{Notations} % title of Table
\vspace{0.5cm}
\centering % used for centering table
\begin{tabular}{|c| c|} % centered columns (4 columns)
\hline % inserts single horizontal line
\hline
\vspace{-0.3cm}
& \\
Fluid velocity & $u_\mu$\\ [1ex]
\hline
\vspace{-0.3cm}
& \\
Membrane velocity & $U_A$\\ [1ex]
\hline
\vspace{-0.3cm}
& \\
Boundary metric & $\eta_{\mu\nu}$\\ [1ex]
\hline
\vspace{-0.3cm}
& \\
Background metric in $(Y^A=\{\rho,y^\mu\})$ & $\bar{G}_{AB}$\\ [1ex]
\hline
\vspace{-0.3cm}
& \\
Background metric in $(X^A=\{r,x^\mu\})$ & $\bar{\cal G}_{AB}$\\ [1ex]
\hline
\vspace{-0.3cm}
& \\
Full metric in $(X^A=\{r,x^\mu\})$ & $ {\cal G}_{AB}$\\ [1ex]
\hline
\vspace{-0.3cm}
& \\
Background metric in arbitrary coordinates & $\bar{\cal W}_{AB}$\\ [1ex]
\hline
\vspace{-0.3cm}
& \\
Full metric in arbitrary coordinates & $ {\cal W}_{AB}$\\ [1ex]
\hline
\vspace{-0.3cm}
& \\
Projector perpendicular to $n_A$ &$\Pi_{AB}=\bar{\cal W}_{AB}-n_A~n_B$\\ [1ex]
\hline
\vspace{-0.3cm}
& \\
Projector perpendicular to both $n_A$ and $U_A$ &$P_{AB}=\bar{\cal W}_{AB}-n_A~n_B+U_A U_B$\\ [1ex]
\hline
\vspace{-0.3cm}
& \\
Projector perpendicular to $u_\mu$ &${\cal P}_{\mu\nu}=\eta_{\mu\nu}+u_\mu u_\nu$\\ [1ex]
\hline
\vspace{-0.3cm}
&\\
Covariant derivative w.r.t background& $\nabla_A$\\
[1ex]
\hline
\vspace{-0.3cm}
&\\
Covariant derivative w.r.t  ${\cal G}_{AB}$& $\bar\nabla_A$\\
[1ex]
\hline
\vspace{-0.3cm}
&\\
~~~~~~~~Covariant derivative w.r.t. induced ~~& ~~~~~~~~~~~~~~~~$\tilde\nabla_\mu$~~~~~~~~~~~~~~~ \\
 metric on the membrane & \\
 [1ex]
\hline
\vspace{-0.3cm}
&\\
~~~~~~~~Covariant derivative projected ~~& ~~~~~~~~~~~~~~~~$\hat\nabla_A$~~~~~~~~~~~~~~~ \\
 along the membrane & ~See equation \eqref{tildedef} for detail definition\\
[1ex]
\hline
\hline
\end{tabular}
\end{table}

\bibliographystyle{JHEP}
\bibliography{larged}

\end{document}